\documentclass[aps,prd,reprint,amsmath,amssymb,superscriptaddress,floatfix,nofootinbib]{revtex4-2}

\usepackage[T1]{fontenc}
\usepackage[utf8]{inputenc}
\usepackage{graphicx}
\usepackage{trimclip}
\usepackage{bm}
\usepackage{placeins}
\usepackage{xcolor}
\usepackage[caption=false,listofformat=subsimple]{subfig} % REVTeX-compatible subfigures (subcaption/caption override REVTeX captions); \subref gives "a", the text supplies the parentheses

\usepackage{hyperref}
\hypersetup{hidelinks}
\begin{document}

\title{ Vacuum Selection by Primordial Helical Magnetic Fields and Domain-Wall Decay }

\author{Jibin He}
 \affiliation{Department of Physics and Chongqing Key Laboratory for Strongly Coupled Physics, Chongqing University, Chongqing 401331, P. R. China}

 \author{Zizhuo Zhao}
\affiliation{
~International Centre for Theoretical Physics Asia-Pacific (ICTP-AP), University of Chinese Academy of Sciences (UCAS), Beijing, China.
}
\affiliation{
~Taiji Laboratory for Gravitational Wave Universe (Beijing/Hangzhou), University of Chinese Academy of Sciences (UCAS), Beijing, China.}

\author{Ligong Bian}
 \email{lgbycl@cqu.edu.cn}
 \affiliation{Department of Physics and Chongqing Key Laboratory for Strongly Coupled Physics, Chongqing University, Chongqing 401331, P. R. China}

\begin{abstract}
We study the effect of helical magnetic fields on domain-wall formation and decay using $(3+1)$-dimensional lattice simulations. In a nonperiodic $Z_2$ model, the electromagnetic source generates an unequal occupation of the two degenerate vacua during wall formation. The formation bias follows $\mathcal B_s\propto\mathcal R_i^{0.78}$ over the range of resistivities studied, where $\mathcal R_i$ characterizes the initial electromagnetic source. Switching the coupling on and off at different times shows that the subsequent wall decay is mainly due to the bias generated during formation. In simulations of a periodic complex field with $N_{\rm DW}=2$, neither resistive magnetohydrodynamics nor vacuum Maxwell dynamics appreciably suppresses the wall area over the simulated interval. We estimate the gravitational-wave spectrum and the abundance of stable scalar particles emitted by the walls, and use the fitted decay timescale to estimate the magnetic-field conditions for wall decay.
\end{abstract}

\maketitle
\section{Introduction}
Long-lived domain walls can dominate the cosmic energy density and spoil standard cosmology~\cite{Zeldovich:1974uw,Kibble:1976sj,Saikawa:2017hiv}. Domain walls can form through the Kibble mechanism~\cite{Kibble:1976sj,Zeldovich:1974uw} when a discrete symmetry is spontaneously broken. In post-inflationary axion models~\cite{Chatterjee:2019rch}, domain walls form and attach to axion strings around the QCD epoch~\cite{Hiramatsu:2012sc,Li:2023gil,Hiramatsu:2010yn,Hiramatsu:2012gg,Lu:2023ayc}. For $N_{\rm DW}>1$, the resulting network can be long-lived. Wall decay can produce gravitational waves in the frequency range of pulsar timing arrays (PTAs)~\cite{Ferreira:2024collapse,Babichev:2023pbf,Blasi:2023sej,Gouttenoire:2023ftk,Bian:2022qbh,Bian:2020urb,Bian:2023dnv} and contribute to the dark matter abundance~\cite{Marsh:2015xka,Li:2023gil,Hiramatsu:2012sc,Hiramatsu:2012gg}.

A potential bias lifts the vacuum degeneracy and exerts a pressure on the walls~\cite{Gelmini:1988sf,Larsson:1996sp}. Other proposals use the Lazarides--Shafi mechanism, in which an enlarged gauge symmetry connects the axion vacua~\cite{Lazarides:1982shafi,Chatterjee:2019rch}. An additional light axion coupled to gluons can reorganize the string--wall network or generate a transient bias~\cite{Lee:2026lightaxion}. Interactions linear in the PQ field can instead align the field thermally before the QCD potential turns on, preventing wall formation~\cite{Zhang:2024imperfect}.

An unequal vacuum occupation can also lead to the disappearance of minority domains even when the bare potential remains degenerate~\cite{Lalak:1993population,Coulson:1996biased,Larsson:1996sp}. Analytic calculations and lattice simulations have examined this population-biased decay~\cite{Hindmarsh:1996scaling,Correia:2014bias,Correia:2018bias,Barbini:2026edx}. Long-range correlations in inflationary initial fluctuations can alter the decay of a biased network~\cite{Gonzalez:2023inflation}. 

We ask how such an occupation imbalance can arise in the early Universe. Here we consider a pseudoscalar field coupled to primordial helical magnetic fields through $\phi F\tilde F$. Primordial magnetic fields may be generated during inflation or phase transitions and subsequently evolve in a conducting plasma~\cite{Turner:1987bw,Vachaspati:1991nm,Durrer:2013pga,Subramanian:2015lua,Di:2025ncl,Di_2021,Di:2024gsl,Yang_2022}. Earlier studies considered the coupled evolution in homogeneous descriptions~\cite{Campanelli:2005ye,Dvornikov:2020hft,Dvornikov:2022cyz}. Lattice simulations with vacuum Maxwell dynamics have found magnetically induced wall decay~\cite{Zhao:2025hzr}, while scalar-field simulations have studied decay from an imposed population bias~\cite{Barbini:2026edx}. It remains to determine how the bias develops in a conducting plasma and how it depends on the vacuum structure.

We first simulate a real scalar field with a nonperiodic double-well potential and study the dependence of vacuum occupation and wall area on the electromagnetic source. By switching the coupling on and off at different times, we test whether wall decay requires continuous electromagnetic forcing. We then compare the nonperiodic model with its periodic complex field counterpart featuring $N_{\rm DW}=2$, using both resistive magnetohydrodynamics (MHD) and vacuum Maxwell dynamics. Using the measured population bias and wall-area evolution, we estimate the gravitational-wave spectrum, dark matter relic abundance and magnetic-field conditions for wall decay.

The simulation framework is given in Sec.~\ref{sec:main_setup}. Section~\ref{sec:main_results} presents the vacuum selection and wall decay results, and Sec.~\ref{sec:main_radiation} discusses gravitational waves and dark matter. The magnetic-field requirements are estimated in Sec.~\ref{sec:mfcons}, including a BBN timing criterion. Section~\ref{sec:discon} concludes.

\section{Simulation framework}\label{sec:main_setup}
We consider an axion-like real pseudoscalar with a nonperiodic $Z_2$-symmetric double-well potential and an electromagnetic coupling,
\begin{equation}
V(\phi)=\frac{m_\phi^2}{8\phi_0^2}(\phi^2-\phi_0^2)^2,
\qquad\mathcal L_{\rm int}=\frac{g_{\phi\gamma\gamma}}4\phi F\tilde F .
\label{eq:V_Z2}
\end{equation}
The two minima at $\pm\phi_0$ have curvature mass $m_\phi$. This real-scalar model has no axion strings. We use conformal time $\tau$, comoving spatial derivatives, and the fields $\mathbf E_c=R^2\mathbf E$ and $\mathbf B_c=R^2\mathbf B$, with $R$ the FLRW scale factor. Primes denote $\partial_\tau$, $\mathcal H=R'/R$, and $g=g_{\phi\gamma\gamma}$. The scalar obeys
\begin{equation}
\phi''+2\mathcal H\phi'-\nabla^2\phi+R^2V_{,\phi}
=-\frac{g}{R^2}\mathbf E_c\cdot\mathbf B_c .
\label{eq:scalar_comoving}
\end{equation}
Neglecting the displacement current in resistive MHD, we use Faraday's law, the modified Ampère equation and Ohm's law,
\begin{align}
\mathbf B_c'&=-\nabla\times\mathbf E_c,\qquad
\nabla\cdot\mathbf B_c=0,\label{eq:faraday}\\
\mathbf J_c&=\nabla\times\mathbf B_c+
g\bigl(\phi'\mathbf B_c+\nabla\phi\times\mathbf E_c\bigr),\label{eq:ampere}\\
\mathbf J_c&=\sigma_c(\mathbf E_c+\mathbf v\times\mathbf B_c),
\label{eq:ohm}
\end{align}
where $\sigma_c=R\sigma$ and $\mathbf J_c$ is the conduction current. Equations~\eqref{eq:ampere} and~\eqref{eq:ohm} determine $\mathbf E_c$ algebraically. Its component parallel to $\mathbf B_c$ sources the scalar. In a fixed helical background this source acts as a local tilt even though the bare potential has degenerate minima. In the simulations the electromagnetic field evolves together with $\phi$.

The fluid has pressure $p=\rho/3$, comoving energy density $\rho_c=R^4\rho$ and velocity $\mathbf v$. Define $D_\tau=\partial_\tau+\mathbf v\cdot\nabla$, $\Theta=\nabla\cdot\mathbf v$ and $Q_c=\boldsymbol{\mathcal J}_c\cdot\mathbf E_c$. The weakly compressible closure is
\begin{align}
\rho_c'&=-\tfrac43\bigl(\mathbf v\cdot\nabla\rho_c+\rho_c\Theta\bigr)+Q_c,\label{eq:fluid_density}\\
D_\tau\mathbf v&=-\frac{\nabla\rho_c}{4\rho_c}
+\frac{3\boldsymbol{\mathcal J}_c\times\mathbf B_c}{4\rho_c}\nonumber\\
&\quad+\frac{\mathbf v}{3}\left(\frac{\mathbf v\cdot\nabla\rho_c}{\rho_c}
+\Theta-\frac{3Q_c}{\rho_c}\right)\nonumber\\
&\quad+\nu_c\left(\nabla^2\mathbf v+\tfrac13\nabla\Theta\right).
\label{eq:fluid_velocity}
\end{align}
Here $\boldsymbol{\mathcal J}_c=\nabla\times\mathbf B_c$ is the total current used in the fluid closure and $\nu_c=1/\sigma_c$ is the comoving kinematic viscosity. The expansion is prescribed. Appendices~\ref{sec:model}--\ref{app:numerics} give the covariant formulation, electric-field inversion and lattice normalization.

The simulations use Dedalus~\cite{2020PhRvR...2b3068B} on a periodic $576^3$ grid with $\tilde L=72$, where $\tilde\phi=\phi/\phi_0$ and $(\tilde\tau,\tilde{\mathbf x})=\omega_*(\tau,\mathbf x)$, with $\omega_*=6R_iH_i$. We initialize the radiation background at $T_i=220$~MeV, $R_i=1$ and $g_*(T_i)\simeq61.3$, using the simulation's Wantz--Shellard fit~\cite{Wantz:2009cosmology}. We take $m_\phi=2\times10^{-10}$~eV, $\phi_0=6\times10^{16}$~GeV and $g_{\phi\gamma\gamma}\phi_0=10^{-3}$, giving $\tilde m_\phi=0.647$. The scalar starts from independent draws in $\tilde\phi\in[-0.1,0.1]$ with zero velocity; the fluid is initially at rest. The spinodal initial state is prescribed at $H_i/m_\phi\simeq0.26$, after the $H=m_\phi$ crossing, and does not reconstruct a thermal phase transition.

We generate the initial divergence-free random magnetic field using transverse and helical Fourier projections~\cite{RoperPol:2022polarization}. All magnetic runs use maximal negative initial helicity ($\mathcal P_{M,i}=-1$), except the controls with maximal positive or zero helicity in Appendix~\ref{app:helicity}. The field carries $1.3\%$ of the radiation energy and peaks at $\tilde k_*=15$. We vary $\tilde\eta=\omega_*/\sigma_c$ from $5\times10^{-6}$ to $5\times10^{-4}$ at fixed initial fields, and vary the coupling, vacuum scale and magnetic parameters in additional runs. Evolution covers $6\le\tilde\tau\le40$, ending at $T\simeq45.5$~MeV with a wall width of $3.1$ cells.

\section{Vacuum selection and domain wall decay}\label{sec:main_results}

\begin{figure}[!htp]
\centering
\subfloat[Population bias\label{fig:B_tau}]{\includegraphics[width=0.8\linewidth]{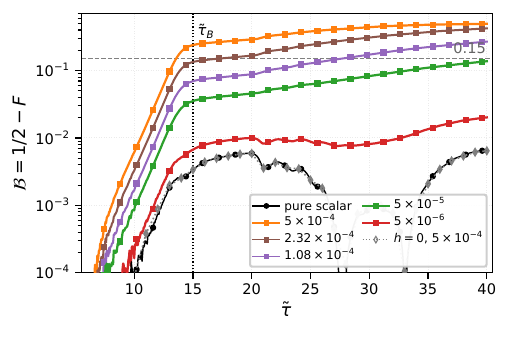}}
\hfill
\subfloat[Population bias at formation\label{fig:deltaBs15_vs_Ri}]{\includegraphics[width=0.8\linewidth]{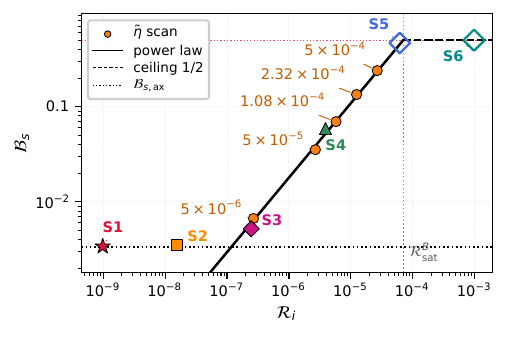}}
\caption{Vacuum selection and its dependence on the initial drive. (a) Population bias; the vertical guide marks $\tilde\tau_B=15$. (b) Formation bias versus $\mathcal R_i$. Circles denote the resistivity scan; other markers denote S1--S6 (Table~\ref{tab:Ri_runs}). Horizontal guides mark the bias in the uncoupled run and complete occupation.}
\label{fig:b_response}
\end{figure}

\subsection{Vacuum selection and initial drive}\label{sec:vacuum_drive}
Let $f_\pm$ denote the volume fractions with $\phi\gtrless0$, $F=\min(f_+,f_-)$ the minority fraction, and $\mathcal B=1/2-F$ the population bias. Before the field settles near its minima, these are sign fractions. Figure~\ref{fig:B_tau} shows that the bias increases with resistivity. The two strongest-drive runs reach $f_+\simeq0.994$ and $0.922$ at $\tilde\tau=40$, while the weakest gives $0.520$, close to the pure-scalar value $0.507$.

At zero initial fluid and scalar velocities, expanding the electric field to leading order in $g_{\phi\gamma\gamma}|\nabla\phi|/\sigma_c$ gives $\mathbf E_c\cdot\mathbf B_c\simeq\mathbf B_c\cdot(\nabla\times\mathbf B_c)/\sigma_c$. For the helical spectra considered here, $2\rho_{B,c}k_{*,c}$ estimates the magnitude of the current helicity and gives the drive scale
\begin{equation}
\mathcal R_i\equiv
\left.\frac{g_{\phi\gamma\gamma}}{\phi_0\omega_*^2}
\frac{2\rho_{B,c}k_{*,c}}{\sigma_c}\right|_i ,
\label{eq:Rcontrol}
\end{equation}
where $\rho_{B,c}=\langle\mathbf B_c^2\rangle/2$ and $k_{*,c}$ is the magnetic spectral peak. $\mathcal R_i$ estimates the source magnitude for the adopted positive coupling; the sign of the initial source is set by the magnetic current helicity. For a Standard Model conductivity $\sigma\sim100T$~\cite{Baym:1997gq,Arnold:2000dr}, holding the other benchmark inputs fixed gives $\mathcal R_i\simeq8\times10^{-22}$, far below the drives simulated here.

We measure $\mathcal B_s=\mathcal B(15)$ by interpolating the histories at $\tilde\tau_B=15$, within the formation transient.
Over $2.68\times10^{-7}\lesssim\mathcal R_i\lesssim2.68\times10^{-5}$, the five primary runs follow
\begin{equation}
\mathcal B_s\simeq884\,\mathcal R_i^{0.782}.
\label{eq:B_response}
\end{equation}
The additional runs cover weaker driving and saturation (Fig.~\ref{fig:deltaBs15_vs_Ri}). The dependence on $\mathcal R_i$ is also tested using
pre-evolved magnetic fields in Appendix~\ref{app:prep_vs_noprep}
(see Fig.~\ref{fig:deltaBs15_prep_measured}). The fit uses the total bias of each run, including its finite-box asymmetry. The pure-scalar control gives $\mathcal B_{s,\rm ax}\simeq0.00335$, which sets the scale below which the magnetic contribution is difficult to distinguish in this realization. By definition, $\mathcal B_s\leq1/2$.

To estimate the network lifetime, we fit the minority fraction with the population-bias template of Ref.~\cite{Barbini:2026edx}. Here $\tau_{\rm ann}$ is the template's characteristic annihilation time, rather than a time of complete wall disappearance. It is distinct from the GW-production endpoint $\tau_{\rm gw}$ and the 1\% area threshold $\tau_d$ used for scalar emission below. Only the two strongest runs have fitted characteristic times within the simulated interval. They agree at the tens-of-percent level with $\tau_{\rm ann}-\tau_B\simeq0.2\tau_B\mathcal B_s^{-0.8}$ at fixed exponent. The other primary lifetimes require extrapolation.

\begin{figure}[!htp]
\centering
\subfloat[Area parameter evolution\label{fig:area_parameter}]{\includegraphics[width=0.8\linewidth]{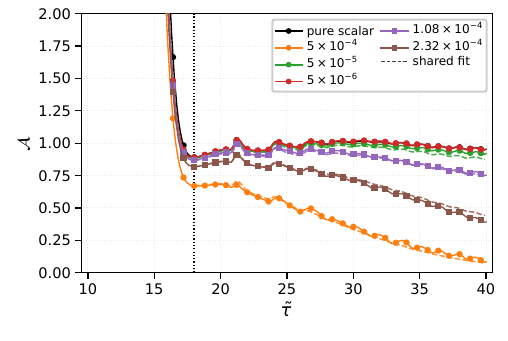}}
\hfill
\subfloat[Initial area deficit\label{fig:deltaA18_vs_Ri}]{\includegraphics[width=0.8\linewidth]{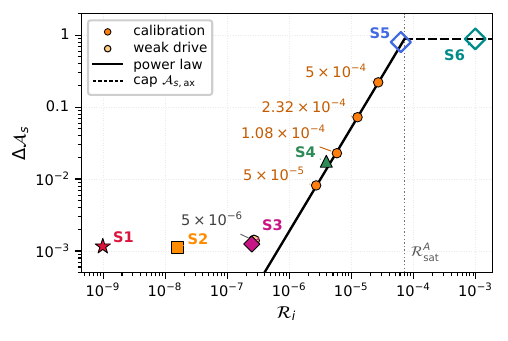}}
\caption{Wall geometry and its dependence on the initial drive. (a) Area histories and the fitted area evolution (dashed); the vertical guide marks $\tilde\tau_A=18$. (b) Initial area deficit versus $\mathcal R_i$. The capped power law is fitted to the four orange circles; the pale circle is the weakest primary run. Other markers denote S1--S6. The saturation intersection is extrapolated.}
\label{fig:a_response}
\end{figure}

The wall area during formation also depends on $\mathcal R_i$, see Fig.~\ref{fig:a_response} for illustration.\footnote{The early bump in Fig.~\ref{fig:area_parameter} is a transient of the sign-change estimator before relaxed walls form, not an enhanced scaling area~\cite{Hiramatsu:2012sc}.} We use the Hubble-normalized area parameter $\mathcal A=A_c/(2\mathcal H V_c)$, where $A_c$ is the comoving wall area and $V_c$ the comoving volume. At $\tilde\tau_A=18$, define $\mathcal A_s=\mathcal A(\tilde\tau_A)$ and $\Delta\mathcal A_s=\mathcal A_{s,\rm ax}-\mathcal A_s$, with the pure-scalar value $\mathcal A_{s,\rm ax}\simeq0.891$. Four primary points give the capped power law
\begin{equation}
\Delta\mathcal A_s\simeq\min\!\bigl(8.422\times10^5\mathcal R_i^{1.440},\mathcal A_{s,\rm ax}\bigr),
\label{eq:deltaA_vs_Ri}
\end{equation}
as shown in Fig.~\ref{fig:deltaA18_vs_Ri}. At weak drive, small differences in the timing of the area turnover affect the deficit measured at $\tilde\tau_A$.

\begin{table*}[!htp]
\centering
\small
\caption{Parameters of the six runs in Fig.~\ref{fig:area_compare_z2_ndw2}. All use a $576^3$ grid, $\tilde L=72$, and $6\le\tilde\tau\lesssim40$. All magnetized runs have maximal negative initial helicity. Within each scalar-field model, the three runs share the scalar potential and initial realization. The electromagnetic inputs of the MHD and vacuum-Maxwell comparisons are matched between scalar-field models. The final column gives the unmasked Hubble-normalized area at $\tilde\tau\simeq40$.}
\label{tab:area_compare_runs}
\setlength{\tabcolsep}{3pt}
\renewcommand{\arraystretch}{1.12}
\begin{tabular}{@{}llcccccc@{}}
\hline
Run & Scalar & Electromagnetism & $\tilde\eta$ & $r_{B,i}$ & $\tilde g$ & $\tilde k_c$ & $\mathcal A(\tilde\tau\simeq40)$ \\
\hline
Pure scalar & $Z_2$ & none & --- & --- & --- & --- & $0.9523$ \\
MHD & $Z_2$ & MHD & $5\times10^{-4}$ & $0.013$ & $10^{-3}$ & $15$ & $0.0835$ \\
Vacuum Maxwell & $Z_2$ & vacuum Maxwell & --- & $0.013$ & $10^{-3}$ & $15$ & $0.0418$ \\
\hline
Pure complex field & $N_{\rm DW}=2$ & none & --- & --- & --- & --- & $0.7764$ \\
MHD & $N_{\rm DW}=2$ & MHD & $5\times10^{-4}$ & $0.013$ & $10^{-3}$ & $15$ & $0.7768$ \\
Vacuum Maxwell & $N_{\rm DW}=2$ & vacuum Maxwell & --- & $0.013$ & $10^{-3}$ & $15$ & $0.7774$ \\
\hline
\end{tabular}
\end{table*}

For $\tilde\tau\ge\tilde\tau_A$, the area histories are reconstructed by
\begin{equation}
\begin{aligned}
\mathcal A(\tilde\tau)&=\mathcal A_{\rm ax}(\tilde\tau)\left[1-\frac{\Delta A_0}{\mathcal A_{s,\rm ax}}\right]^{G(\tilde\tau)},\\
G(\tilde\tau)&=1+\kappa_A(\tilde\tau-\tilde\tau_A)^{p_A}.
\end{aligned}
\label{eq:A_from_deltaA}
\end{equation}
where $\mathcal A_{\rm ax}$ is the pure-scalar history and $\mathcal A_{s,\rm ax}=\mathcal A_{\rm ax}(\tilde\tau_A)$. For the dashed curves in Fig.~\ref{fig:area_parameter}, $\Delta A_0$ is set by Eq.~\eqref{eq:deltaA_vs_Ri}, while $\kappa_A\simeq0.0444$ and $p_A\simeq1.69$ are shared by all five primary runs. The joint fit over $18\le\tilde\tau\le40$ has an rms area residual of $0.016$. Only the initial deficit changes between runs, with $0\le\mathcal A\le\mathcal A_{\rm ax}$. Appendix~\ref{app:area_response} gives the fitting procedure and extrapolation limits.

The coupling controls in Appendix~\ref{app:supp_figs} show that the later area loss is governed predominantly by the vacuum imbalance established during formation.

\subsection{Persistence of the periodic $N_{\rm DW}=2$ wall network}
\label{sec:periodic_limit}
\begin{figure}[!htp]
\centering
\subfloat[Nonperiodic $Z_2$ scalar\label{fig:z2_area_compare}]{\includegraphics[width=0.8\linewidth]{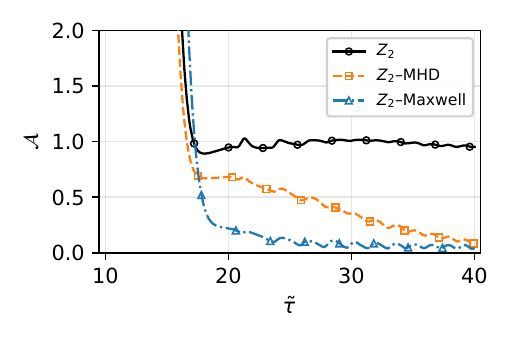}}
\hfill
\subfloat[Periodic complex field, $N_{\rm DW}=2$\label{fig:ndw2_area_three}]{\includegraphics[width=0.8\linewidth]{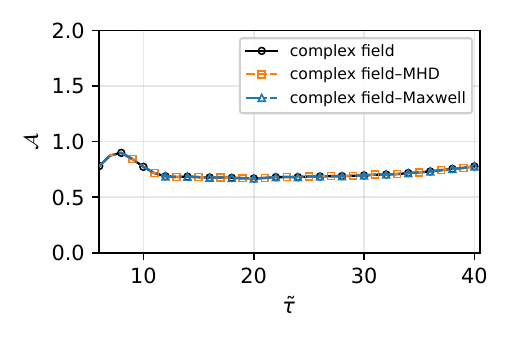}}
\caption{Evolution of the area parameter for (a) the nonperiodic $Z_2$ scalar and (b) the periodic $N_{\rm DW}=2$ complex field. Circles, squares and triangles distinguish the pure-field, MHD-coupled and Maxwell-coupled runs. Parameters and final areas are listed in Table~\ref{tab:area_compare_runs}.}
\label{fig:area_compare_z2_ndw2}
\end{figure}

To examine the effect of periodicity, we simulate a complex field $\Phi=\phi_1+i\phi_2$ with a periodic $N_{\rm DW}=2$ potential, using the same initial scalar field in the uncoupled, MHD and vacuum Maxwell runs. All three complex-field runs use the $576^3$ grid, $\tilde L=72$ and $6\le\tilde\tau\lesssim40$. The angular mass grows with cooling and reaches $\tilde m_a\simeq0.323$. The MHD run uses $\tilde\eta=5\times10^{-4}$, $\tilde g=f_ag_{\phi\gamma\gamma}=10^{-3}$, and an initial magnetic energy fraction $r_{B,i}=0.013$ with maximal negative helicity and spectral peak $\tilde k_c=15$. These electromagnetic inputs match those of the $Z_2$--MHD comparison, including the initial magnetic-field realization. Table~\ref{tab:area_compare_runs} lists the runs in Fig.~\ref{fig:area_compare_z2_ndw2}. The vacuum Maxwell run uses the same magnetic field and coupling as its $Z_2$ counterpart. Appendix~\ref{app:ndw2} gives the complex-field model and phase distributions.

Figure~\ref{fig:area_compare_z2_ndw2} compares the area evolution in the two models. In the $Z_2$ model both MHD and vacuum Maxwell dynamics strongly suppress the late wall area. In the periodic model, the area histories of the uncoupled, MHD-coupled and vacuum Maxwell runs remain close. The final area parameters of the three complex-field runs differ by less than $0.2\%$. See Appendix~\ref{app:ndw2} for the histograms of the vacuum distribution, which indicate no appreciable additional population bias signature even at the final time of the simulations.

\begin{figure}[!htp]
\centering
\includegraphics[width=0.8\linewidth]{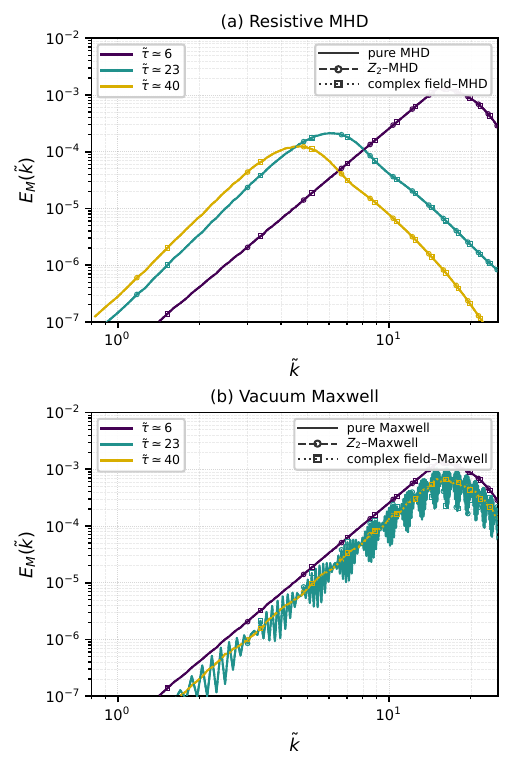}
\caption{Magnetic energy spectra for (a) resistive MHD and (b) vacuum Maxwell dynamics, for the matched inputs in Table~\ref{tab:area_compare_runs}. Colors distinguish $\tilde\tau\simeq6$, $23$, and $40$; line and marker styles distinguish the pure-electromagnetic, $Z_2$-coupled and complex-field-coupled runs.}
\label{fig:magnetic_sixrun}
\end{figure}

Figure~\ref{fig:magnetic_sixrun} compares pure MHD and pure vacuum Maxwell dynamics with their $Z_2$ and periodic complex-field counterparts at three times. Here $E_M(\tilde k)$ is the magnetic energy density per unit lattice wavenumber (Appendix~\ref{app:mhd}). The spectra in the scalar-coupled runs closely follow those obtained without the scalar coupling. The MHD spectra move toward smaller wavenumbers, whereas the Maxwell spectra retain a peak near the initial magnetic scale. The different wall-area responses therefore occur without comparably large changes in the magnetic field spectra.

\begin{figure}[!htp]
\centering
\includegraphics[width=0.8\linewidth]{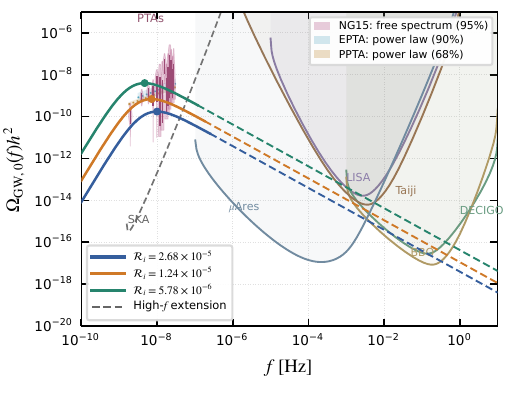}
\caption{GW template spectra for the three largest drives at the simulation vacuum scale. This vacuum scale overproduces dark matter under the stable-scalar emission assumptions of Fig.~\ref{fig:scalar_wall_abundance}. Dashed tails above $f/f_p=25$ continue the analytic template without an ultraviolet turnover. PTA intervals and projected sensitivities use the assumptions described in Appendix~\ref{sm:observational_context}.}
\label{fig:gw_population}
\end{figure}

\section{Gravitational waves and dark matter relic abundance}\label{sec:main_radiation}
We use the population-bias template of Ref.~\cite{Barbini:2026edx} to estimate the gravitational-wave spectrum from the measured $\mathcal B_s$. For the three largest drives in descending order, Fig.~\ref{fig:gw_population} gives peak frequencies $f_p=(10.0,7.22,4.72)$~nHz and peak energy-density fractions $\Omega_p h^2=(1.74\times10^{-10},6.90\times10^{-10},3.98\times10^{-9})$. The template uses the benchmark wall tension $\sigma_{\rm w}=2m_\phi\phi_0^2/3$ and the emission endpoint $\tau_{\rm gw}-\tau_B=2.85(\tau_{\rm ann}-\tau_B)$, with the factor calibrated in Ref.~\cite{Barbini:2026edx}. All three endpoints extend beyond our simulated interval. At fixed mass and vacuum scale, smaller drives give later decay and a higher estimated peak amplitude. The estimated peaks lie in the PTA frequency band. The comparison in Fig.~\ref{fig:gw_population} does not constrain the wall contribution separately from other sources of the background.

\begin{figure}[!htp]
\centering
\includegraphics[width=0.8\linewidth]{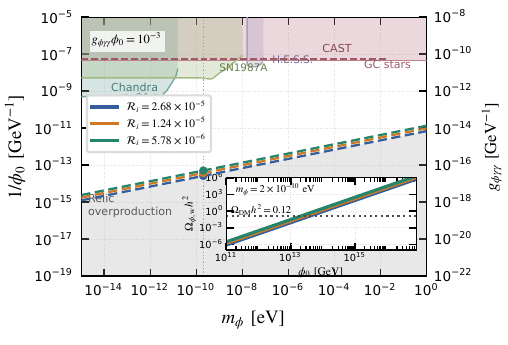}
\caption{Stable-scalar abundance for $b_\phi=1$ and $\epsilon_{\rm em}=1.1$. Dashed boundaries assume radiation-era similarity; dots mark the simulated mass. Gray shading overproduces dark matter for all three drives under this extrapolation. Photon limits use $g_{\phi\gamma\gamma}\phi_0=10^{-3}$. Inset: vacuum-scale dependence at the simulated mass. Assumptions and sources are in Appendix~\ref{sm:observational_context}.}
\label{fig:scalar_wall_abundance}
\end{figure}

We estimate the stable-scalar relic abundance from Eq.~\eqref{eq:A_from_deltaA} using a wall-energy-release prescription~\cite{Zhao:2025hzr}, with the pure-scalar baseline held at its $\tilde\tau_A$ value for the area extrapolation (Appendix~\ref{app:scalar_relic}). We integrate the release from $\tilde\tau_A=18$ until the reconstructed area reaches $1\%$ of its reference value, a separate endpoint for scalar production. Figure~\ref{fig:scalar_wall_abundance} uses a scalar energy fraction $b_\phi=1$ and a mean emission energy $1.1m_\phi$. The relic density scales as $\phi_0^2$ at fixed dimensionless response, while the template GW amplitude scales as $\phi_0^4$. At the simulated mass, the observed dark-matter density bounds the vacuum scale by $\phi_0\lesssim(3.83,2.69,1.97)\times10^{13}$~GeV for the same three drives. Saturating the dark-matter bound at fixed dimensionless response gives peak amplitudes $\Omega_p h^2\simeq(2.9,2.8,4.6)\times10^{-23}$. At these amplitudes, the wall signal would be a negligible contribution to the measured PTA background. A similar tension between the PTA signal and the dark matter abundance was found by Zhao et al.~\cite{Zhao:2025hzr} using vacuum Maxwell dynamics.

\section{Magnetic field constraints}\label{sec:mfcons}
Let $r_{B,i}=\rho_{B,i}/\rho_{\rm rad,i}$. We define the initial physical magnetic scale as $\ell_i=R_i/k_{*,c}$, using the spectral peak rather than the integral correlation length. For the benchmark helical spectral shape, the drive at fixed conductivity and scalar parameters is proportional to $r_{B,i}/(H_i\ell_i)$. At $\tilde\eta=5\times10^{-4}$,
\begin{equation}
\mathcal R_i\simeq2.68\times10^{-5}
\left(\frac{r_{B,i}}{0.013}\right)
\left(\frac{1}{90H_i\ell_i}\right).
\label{eq:magnetic_drive_slice}
\end{equation}

\begin{figure}[!htp]
\centering \includegraphics[width=0.8\linewidth]{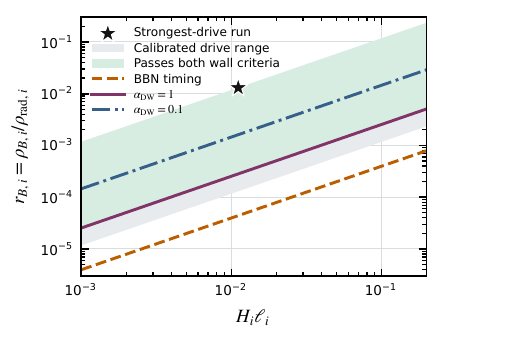}
\caption{Initial magnetic energy fraction and coherence scale at $\tilde\eta=5\times10^{-4}$ and the benchmark scalar parameters. Curves mark the BBN timing criterion and wall-energy fractions $\alpha_{\rm DW}=1$ and $0.1$. Green shading identifies drives within the fitted range satisfying the BBN timing and conditions for avoiding wall domination.} \label{fig:magnetic_parameter_space}
\end{figure}

Figure~\ref{fig:magnetic_parameter_space} shows the resulting constraints on $r_{B,i}$ and $H_i\ell_i$. At $H_i\ell_i=1/90$, requiring that wall-sourced GW emission end by $T=1$~MeV gives $r_{B,i}\gtrsim4.4\times10^{-5}$. This BBN timing criterion corresponds to a drive below the range of our fit. The wall fraction $\alpha_{\rm DW}=\rho_{\rm DW}/\rho_{\rm tot}$ is evaluated at $\tau_{\rm ann}$ using the pure-scalar scaling area and the benchmark tension. Requiring $\alpha_{\rm DW}<1$ gives $r_{B,i}\gtrsim2.8\times10^{-4}$, while $\alpha_{\rm DW}<0.1$ requires $r_{B,i}\gtrsim1.6\times10^{-3}$. These estimates assume that the cosmological magnetic contribution is given by the excess bias measured in the simulations, and extrapolate the lifetime template (Appendix~\ref{sec:cosmology}). We assume the same relation between $\mathcal R_i$ and the bias at other coherence scales.

\section{Discussion and conclusion}\label{sec:discon}
In the nonperiodic $Z_2$ model, helical magnetic fields generate a population bias during domain-wall formation. The subsequent decay is mainly due to this bias rather than continuous electromagnetic forcing. At fixed magnetic energy density, spectral shape and scalar parameters, a smaller magnetic coherence scale or a lower conductivity increases the initial source $\mathcal R_i$. At fixed magnetic and background parameters, $\mathcal R_i\propto g_{\phi\gamma\gamma}$ for fixed $\phi_0$, and $\mathcal R_i\propto\phi_0^{-2}$ for fixed $g_{\phi\gamma\gamma}\phi_0$. The source estimated at the Standard Model conductivity is much smaller than in our simulations. Lower conductivities may arise in a dark plasma or during incomplete thermalization~\cite{Brandenburg:2023cao,Allahverdi:2020bys}, provided that the MHD approximation remains applicable. The same mechanism may generate unequal vacuum occupation in nonperiodic potentials with more than two degenerate vacua, although the quantitative relations obtained here are specific to the simulated $Z_2$ model and initial conditions.
For the nonperiodic model, the estimated gravitational-wave peaks lie in the PTA frequency band. Under the adopted stable-scalar emission prescription, however, the dark matter abundance bound reduces the peak amplitudes to a few times $10^{-23}$ in $\Omega_p h^2$, well below current PTA sensitivity.

The periodic complex field behaves differently. For $N_{\rm DW}=2$, neither resistive MHD nor vacuum Maxwell dynamics appreciably suppresses the wall area over the simulated interval. As discussed in Appendix~\ref{app:ndw2},
the electromagnetic coupling preserves the discrete shift symmetry and cannot systematically favor a particular vacuum in a symmetric initial ensemble. This argument applies to arbitrary $N_{\rm DW}>1$ in the class of models considered here, although it does not exclude other effects on string--wall dynamics. Simulations with $N_{\rm DW}>2$ are needed to test the corresponding network evolution quantitatively.

Our simulations start at $T_i=220$~MeV and do not follow the full evolution from PQ symmetry breaking to the QCD epoch. The radial-to-angular mass hierarchy is also limited in the complex-field simulations. The evolution of the string--wall network through the full PQ--QCD history remains to be studied. Magnetic fields generated during inflation or the electroweak phase transition may have different spectra and coupling histories at the QCD epoch. If the electromagnetic coupling is effective at an earlier epoch, even a weaker source may appreciably alter the scalar distribution or velocity before wall formation. The resulting population bias in a nonperiodic potential may therefore depend on the earlier evolution of both fields. 

\begin{acknowledgments}

This work was supported in part by the National Natural Science Foundation of China under Grant Nos. 12675130, 12322505, 12547101, and Fundamental Research Funds for the Central Universities Grant No. 2026CDJZKPT-10.
L.B. also acknowledges Chongqing Natural Science Foundation under Grant No. CSTB2024NSCQ-JQX0022.
\end{acknowledgments}

\FloatBarrier
\appendix
\section{Model details}
\label{sec:model}

\subsection{Covariant formulation}

We consider a spatially flat Friedmann--Lema\^{i}tre--Robertson--Walker (FLRW) universe with metric signature $(+---)$ and conformal time $\tau$. The line element is
\begin{equation}
ds^2=R^2(\tau)(d\tau^2-\delta_{ij}dx^idx^j),
\end{equation}
where $R(\tau)$ is the scale factor.

The scalar and gauge-field dynamics, with the plasma represented by a current $J^\mu$, follow from the action
\begin{align}
\label{eq:S}
S=&\int d^4x\sqrt{-g}\Biggl[
\frac{1}{2}\partial_\mu\phi\partial^\mu\phi
-V(\phi)
-\frac{1}{4}F_{\mu\nu}F^{\mu\nu}\nonumber\\
&-A_\mu J^\mu
+\frac{g_{\phi\gamma\gamma}}{4}\phi F^{\mu\nu}\tilde{F}_{\mu\nu}
\Biggr],
\end{align}

Here $g$ is the determinant of the metric, $F_{\mu\nu}=\partial_\mu A_\nu-\partial_\nu A_\mu$ is the field strength, $\tilde{F}^{\mu\nu}=\frac{1}{2\sqrt{-g}}\epsilon^{\mu\nu\alpha\beta}F_{\alpha\beta}$ is its dual, and $J^\mu$ is the plasma four-current. The axion-photon coupling is $g_{\phi\gamma\gamma}$.

The $Z_2$ model uses a symmetric double-well potential with vacua at $\phi=\pm\phi_0$,
\begin{equation}
\label{eq:V_Z2_app}
V(\phi)
=
\frac{\lambda}{4}(\phi^2-\phi_0^2)^2,
\qquad
\lambda=\frac{m_{\phi}^2}{2\phi_0^2}.
\end{equation}
Here $m_\phi$ is the curvature mass at either vacuum; the numerical values of $m_\phi$ and $\phi_0$ are specified in Sec.~\ref{sec:dimless}.

Walls develop from small random fluctuations near the hilltop, with vanishing initial scalar velocity. Their unstable growth is Kibble-like (spinodal) in character \cite{Kibble:1976sj}. We study how the helical magnetic field affects vacuum selection during this growth and how the resulting asymmetry influences the wall network.

The potential is invariant under $\phi\to-\phi$. In a fixed helical electromagnetic background, however, the coupling term $\phi F\tilde F$ is not separately invariant, because the prescribed helical background is not transformed with $\phi$. For the initial spectra considered here, helicity fixes the sign of the volume-averaged source $\langle\mathbf E\cdot\mathbf B\rangle$, while its local value can vary across the box. This coupling can therefore act as an effective tilt between the two scalar vacua. For a prescribed electromagnetic background, the scalar equation can be written with $V_{\rm eff}(\phi;\mathbf{x},t)=V(\phi)+g_{\phi\gamma\gamma}\phi\,\mathbf E\cdot\mathbf B$ in our sign convention. This is a local, time-dependent description of the source, rather than an added constant splitting of the bare potential; the electromagnetic fields evolve self-consistently in the simulations.

Varying the action with respect to $A_\nu$ and projecting onto the electric and magnetic fields measured by a comoving observer yields the modified Maxwell equations in conformal coordinates,
\begin{align}
\nabla\cdot\mathbf{E}
=&R\rho_e+
g_{\phi\gamma\gamma}
\mathbf{B}\cdot\nabla\phi,
\\
\nabla\times\mathbf{B}
=&R\,\mathbf{J}
+\frac{1}{R^2}\partial_\tau(R^2\mathbf{E})
\nonumber\\
&-
g_{\phi\gamma\gamma}
\left(
\phi'\mathbf{B}
-\mathbf{E}\times\nabla\phi
\right),
\end{align}
where $\rho_e$ and $\mathbf{J}$ are the physical charge and current densities (the coordinate components of the four-current are $j^0=\rho_e/R$ and $j^i=J^i/R$), and a prime denotes a derivative with respect to $\tau$.

\subsection{Comoving pseudoscalar--MHD system}
\label{sec:comoving}

We use the dimensional comoving variables $\mathbf{B}_c=R^2\mathbf{B}$, $\mathbf{E}_c=R^2\mathbf{E}$, $\mathbf{J}_c=R^3\mathbf{J}$, and $\sigma_c=R\sigma$.
The comoving energy density is $R^4\rho$, and the velocity $\mathbf v$ is unchanged.

The generalized Ohm's law in comoving coordinates is
\begin{equation}
\mathbf{J}_c
=
\sigma_c
(\mathbf{E}_c
+\mathbf{v}\times\mathbf{B}_c),
\end{equation}
where $\mathbf{J}_c$ is the comoving conduction current. For the conductivities considered here the displacement current is small compared with the conduction current (resistive MHD approximation), and the modified Amp\`{e}re equation can be inverted for the electric field,
\begin{align}
\mathbf{E}_c
=&\frac{1}{\sigma_c^2+g_{\phi\gamma\gamma}^2|\nabla\phi|^2}
\biggl[\sigma_c\mathbf{K}
\nonumber\\
&\quad+g_{\phi\gamma\gamma}(\nabla\phi\times\mathbf{K})
+\frac{g_{\phi\gamma\gamma}^2}{\sigma_c}
(\mathbf{K}\cdot\nabla\phi)\nabla\phi\biggr],
\end{align}
where
\begin{equation}
\mathbf{K}
\equiv
\nabla\times\mathbf{B}_c
-\sigma_c\mathbf{v}\times\mathbf{B}_c
+
g_{\phi\gamma\gamma}\phi'\mathbf{B}_c.
\end{equation}

The magnetic field then obeys the comoving Faraday equation,
\begin{equation}
\partial_\tau\mathbf{B}_c
=
-\nabla\times\mathbf{E}_c.
\end{equation}

The scalar equation in the same comoving coordinates is
\begin{align}
\phi''={}&-2\mathcal H\phi'+\nabla^2\phi-R^2V_{,\phi}
\nonumber\\
&-\frac{g_{\phi\gamma\gamma}}{R^2}\,
\mathbf E_c\cdot\mathbf B_c,
\label{eq:scalar_comoving_app}
\end{align}
where $\mathcal H=R'/R$, $V_{,\phi}=\lambda\phi(\phi^2-\phi_0^2)$, and spatial derivatives are with respect to the comoving position.

For the radiation-like fluid, $p=\rho/3$, define the comoving energy density $\rho_c=R^4\rho$ and the total current $\boldsymbol{\mathcal J}_c=\nabla\times\mathbf B_c$. This total current, used in the effective fluid closure, is distinct from the conduction current $\mathbf J_c$ in Ohm's law. In the weakly compressible approximation, the fluid equations are
\begin{align}
\partial_\tau\rho_c
={}&-\frac43\left(\mathbf v\cdot\nabla\rho_c
+\rho_c\nabla\cdot\mathbf v\right)
+\boldsymbol{\mathcal J}_c\cdot\mathbf E_c,
\label{eq:fluid_energy_comoving}\\
\partial_\tau\mathbf v
={}&-\mathbf v\cdot\nabla\mathbf v
-\frac{\nabla\rho_c}{4\rho_c}
+\frac{3\boldsymbol{\mathcal J}_c\times\mathbf B_c}{4\rho_c}
\nonumber\\
&+\mathbf v\left[\frac13\left(
\frac{\mathbf v\cdot\nabla\rho_c}{\rho_c}
+\nabla\cdot\mathbf v\right)
-\frac{\boldsymbol{\mathcal J}_c\cdot\mathbf E_c}{\rho_c}\right]
\nonumber\\
&+\nu_c\left[\nabla^2\mathbf v
+\frac13\nabla(\nabla\cdot\mathbf v)\right],
\label{eq:fluid_velocity_comoving}
\end{align}
where $\nu_c$ is the comoving kinematic viscosity. We set $\nu_c=\eta_c=1/\sigma_c$, corresponding to a magnetic Prandtl number of unity. The fluid is initialized with the background radiation density, and the expansion is prescribed independently through $R(\tau)$.

\section{Simulation setup, numerical formulation and resolution}
\label{app:numerics}
\subsection{Dimensionless variables and simulation parameters}
\label{sec:dimless}

We initialize the simulations at $T_i=220~\mathrm{MeV}$ and $R_i=1$. The primary scan uses $\phi_0=6\times10^{19}~\mathrm{MeV}$ and a constant curvature mass $m_\phi=2\times10^{-16}~\mathrm{MeV}$. These values provide a QCD-axion-motivated mass scale for a decay constant comparable to $\phi_0$ \cite{Wantz:2009susceptibility,GrillidiCortona:2015mass}, place the initial roll and oscillations within the simulated epoch, and keep the walls sufficiently broad for a resolved evolution. The expansion follows the Wantz--Shellard background \cite{Wantz:2009cosmology}, with $g_*(T_i)\simeq61.3$; the scalar mass itself is not varied with temperature.

We express the numerical setup in units set by $f_*=\phi_0$ and $\omega_*=6R_iH_i$. The dimensionless field and coordinates are
\begin{equation}
\tilde\phi=\frac{\phi}{\phi_0},\qquad
\tilde\tau=\omega_*\tau,\qquad
\tilde{\mathbf x}=\omega_*\mathbf x.
\label{eq:simulation_units}
\end{equation}
The corresponding parameters are $\tilde m_\phi=m_\phi/\omega_*=0.647$, $\tilde g=\phi_0g_{\phi\gamma\gamma}=10^{-3}$ and $\tilde\eta=\omega_*/\sigma_c=1/\tilde\sigma$, where $\sigma_c=R\sigma$. Thus the physical coupling is $g_{\phi\gamma\gamma}\simeq1.67\times10^{-23}~\mathrm{MeV}^{-1}$. Electromagnetic fields and fluid energy are normalized to $w^2=\rho_{{\rm rad},i}$, so that $\tilde\rho_B=R^4\rho_B/\rho_{{\rm rad},i}$. The full conversion rules are given in Sec.~\ref{app:dimensionless}.

The primary scan varies $\tilde\eta$ over $5\times10^{-4}$, $2.32\times10^{-4}$, $1.08\times10^{-4}$, $5\times10^{-5}$ and $5\times10^{-6}$ while holding the initial scalar and magnetic fields fixed. The initial field has maximal negative helicity, $\tilde\rho_{B,i}=0.013$ and $\tilde k_{*,i}=k_c=15$, as specified in Sec.~\ref{sec:main_setup}. Additional runs vary the coupling, vacuum scale, magnetic energy and spectral peak to test the dependence on $\mathcal R_i$; their inputs are listed in Table~\ref{tab:Ri_runs}. The accessible $k_c$ range is limited by high-wavenumber resolution and the need to keep the evolving peak above $\tilde k_{\min}=2\pi/\tilde L$; extending it while resolving the walls requires a larger grid.

At each grid point, $\tilde\phi$ is drawn independently from $[-0.1,0.1]$, with vanishing initial scalar velocity. The fluid starts at rest with $\ln\tilde\rho=0$. The main scan uses the magnetic field directly, without prior MHD evolution; pre-evolved initial data are compared in Sec.~\ref{app:prep_vs_noprep}. The Fourier projection of the scalar initial field is described in Sec.~\ref{app:dimensionless}.

We solve the coupled system with Dedalus \cite{2020PhRvR...2b3068B} on a periodic $576^3$ grid, with dimensionless box length $\tilde L=72$ and spacing $\Delta\tilde x=0.125$. This box has comoving length $12(R_iH_i)^{-1}$. Evolution runs from $\tilde\tau_i=6$ to $\tilde\tau_f=40$, ending at $R\simeq7.90$ and $T\simeq45.5$~MeV. The wall width $\tilde\delta_{\rm wall}=2/(R\tilde m_\phi)$ still spans $3.1$ cells at the endpoint. The time integration and resolution criteria are specified in Sec.~\ref{app:lattice_resolution}.

\subsection{Normalization and parameter conversion}
\label{app:dimensionless}

The field and coordinate units are defined in Sec.~\ref{sec:dimless}.  Define $w^2=\rho_{{\rm rad},i}=3H_i^2/(8\pi G)$. The electromagnetic fields, fluid density and transport coefficients obey
\begin{equation}
\begin{gathered}
\tilde{\mathbf B}=\frac{\mathbf B_c}{w},\qquad
\tilde{\mathbf E}=\frac{\mathbf E_c}{w},\qquad
\tilde\rho=\frac{\rho_c}{w^2},\\
\tilde\sigma=\frac{\sigma_c}{\omega_*},\qquad
\tilde\eta=\omega_*\eta_c=\frac{1}{\tilde\sigma},\qquad
\tilde\nu=\omega_*\nu_c=\tilde\eta.
\label{eq:lattice_conversion}
\end{gathered}
\end{equation}
Here $\mathbf B_c=R^2\mathbf B$, $\mathbf E_c=R^2\mathbf E$ and $\rho_c=R^4\rho$. The peculiar velocity $\mathbf v$ and scale factor $R$ are already dimensionless and are not rescaled. In the lattice equations and field histograms, $\phi$ is shorthand for $\tilde\phi=\phi_{\rm phys}/\phi_0$, with vacua at $\pm1$; spatial derivatives act on $\tilde{\mathbf x}$ and primes denote derivatives with respect to $\tilde\tau$.

The scalar source coefficient differs from the dimensionless electromagnetic coupling $\tilde g=\phi_0g_{\phi\gamma\gamma}$. It is $\hat g=\tilde g\,w^2/(\phi_0^2\omega_*^2)$, approximately $0.137$ for the primary benchmark. The initial-drive parameter of the main text therefore becomes
\begin{equation}
\mathcal R=\hat g\,\tilde\eta\,(2\tilde\rho_B\tilde k_*),
\label{eq:R_lattice}
\end{equation}
with $\tilde k_*=k_{*,c}/\omega_*$. At the fixed initial background, $\hat g=0.137263(\tilde g/10^{-3})(6\times10^{19}~\mathrm{MeV}/\phi_0)^2$. Table~\ref{tab:Ri_runs} records both coupling parameters and the resulting $\mathcal R_i$ for each run.

The common comparison times are $\tilde\tau_B=15$ for vacuum occupation and $\tilde\tau_A=18$ for the area deficit. Their comoving values are $\tau_B=5/(2R_iH_i)$ and $\tau_A=3/(R_iH_i)$; the endpoints $\tilde\tau_i=6$ and $\tilde\tau_f=40$ correspond to $\tau_i=(R_iH_i)^{-1}$ and $\tau_f=20/(3R_iH_i)$. The area parameter is invariant under this change of units: $\mathcal A=A_c/(2\mathcal H V_c)=\tilde A_c/(2\tilde{\mathcal H}\tilde V_c)$. For comparison with conformal-time-normalized quantities in the literature, $\mathcal A_\tau=(\tau/2)A_c/V_c=(\mathcal H\tau)\mathcal A$; all area histories and fits in this work use $\mathcal A$.

The initial scalar draws specified in Sec.~\ref{sec:dimless} are projected onto the retained real-Fourier modes. Removing the grid Nyquist components gives a low-pass projection at the grid cutoff. The $3/2$ padding used for nonlinear evaluation adds no separate smoothing window.

\begin{table*}[!t]
\centering
\small
\caption{Parameters and formation-stage biases for the runs in Fig.~\ref{fig:deltaBs15_vs_Ri}. The five resistivity-scan points determine the bias fit; S1--S6 are additional tests. $\phi_0$ is in MeV. The secondary column gives $\Delta\mathcal B_s=\mathcal B_s-0.003347$.}
\label{tab:Ri_runs}
\setlength{\tabcolsep}{1.7pt}
\renewcommand{\arraystretch}{1.12}
\begin{tabular}{@{}l*{9}{c}p{24mm}@{}}
\hline
Run & $\tilde\eta$ & $\tilde g$ & $\tilde\rho_{B,i}$ & $k_c$ & $\phi_0$ [MeV] & $\hat g$ & $\mathcal{R}_i$ & $\mathcal{B}_s$ & $\Delta\mathcal{B}_s$ & Role \\
\hline
$\tilde\eta$-scan & $5\times10^{-6}$ & $10^{-3}$ & $0.013$ & $15$ & $6\times10^{19}$ & $0.137$ & $2.68\times10^{-7}$ & $0.00668$ & $0.00333$ & Fit \\
$\tilde\eta$-scan & $5\times10^{-5}$ & $10^{-3}$ & $0.013$ & $15$ & $6\times10^{19}$ & $0.137$ & $2.68\times10^{-6}$ & $0.0353$ & $0.0320$ & Fit \\
$\tilde\eta$-scan & $1.08\times10^{-4}$ & $10^{-3}$ & $0.013$ & $15$ & $6\times10^{19}$ & $0.137$ & $5.78\times10^{-6}$ & $0.0696$ & $0.0663$ & Fit \\
$\tilde\eta$-scan & $2.32\times10^{-4}$ & $10^{-3}$ & $0.013$ & $15$ & $6\times10^{19}$ & $0.137$ & $1.24\times10^{-5}$ & $0.134$ & $0.131$ & Fit \\
$\tilde\eta$-scan & $5\times10^{-4}$ & $10^{-3}$ & $0.013$ & $15$ & $6\times10^{19}$ & $0.137$ & $2.68\times10^{-5}$ & $0.239$ & $0.236$ & Fit \\
\hline
S1 & $8.50\times10^{-6}$ & $2.01\times10^{-2}$ & $0.0118$ & $17.6$ & $5.97\times10^{21}$ & $2.78\times10^{-4}$ & $9.80\times10^{-10}$ & $0.00340$ & $5.09\times10^{-5}$ & \shortstack[l]{Fixed\\background\\test} \\
S2 & $7.62\times10^{-4}$ & $1.06\times10^{-6}$ & $2.21\times10^{-4}$ & $17.6$ & $1.42\times10^{19}$ & $2.63\times10^{-3}$ & $1.55\times10^{-8}$ & $0.00351$ & $1.62\times10^{-4}$ & \shortstack[l]{Weak\\response} \\
S3 & $7.96\times10^{-4}$ & $2.17\times10^{-3}$ & $1.31\times10^{-6}$ & $18.6$ & $1.30\times10^{19}$ & $6.37$ & $2.46\times10^{-7}$ & $0.00520$ & $1.85\times10^{-3}$ & \shortstack[l]{Weak\\response} \\
S4 & $5.24\times10^{-6}$ & $2.11\times10^{-2}$ & $6.34\times10^{-6}$ & $16.8$ & $1.73\times10^{18}$ & $3.49\times10^{3}$ & $3.90\times10^{-6}$ & $0.0589$ & $0.0556$ & \shortstack[l]{Intermediate\\response} \\
S5 & $2.80\times10^{-5}$ & $5.93\times10^{-6}$ & $6.64\times10^{-3}$ & $21.7$ & $6.18\times10^{17}$ & $7.67$ & $6.19\times10^{-5}$ & $0.467$ & $0.463$ & \shortstack[l]{Near\\saturation} \\
S6 & $1.00\times10^{-3}$ & $1.75\times10^{-2}$ & $4.60\times10^{-3}$ & $19.1$ & $3.95\times10^{19}$ & $5.54$ & $9.80\times10^{-4}$ & $0.500$ & $0.497$ & \shortstack[l]{No wall\\formation} \\
\hline
\end{tabular}
\end{table*}

\subsection{Lattice equations, background and resolution}
\label{app:lattice_resolution}

\begingroup
\setlength{\abovedisplayskip}{4pt}
\setlength{\belowdisplayskip}{4pt}
\setlength{\abovedisplayshortskip}{2pt}
\setlength{\belowdisplayshortskip}{2pt}
\setlength{\jot}{1pt}
\allowdisplaybreaks[1]

The lattice form of Eqs.~\eqref{eq:scalar_comoving_app}--\eqref{eq:fluid_velocity_comoving}, together with Faraday's law, is
\begin{align}
\phi''
&=
-2\tilde{\mathcal H}\phi'
+\tilde\nabla^2\phi
-R^2\frac{1}{2}\tilde m_\phi^2\,\phi(\phi^2-1)
\nonumber\\
&\quad
-\frac{\hat g}{R^2}\,\tilde{\mathbf E}\cdot\tilde{\mathbf B},
\\
\frac{\partial\tilde{\mathbf B}}{\partial\tilde\tau}
&=
-\tilde\nabla\times\tilde{\mathbf E},
\\
\frac{\partial\ln\tilde\rho}{\partial\tilde\tau}
&=
-\frac43
\left(
\mathbf v\cdot\tilde\nabla\ln\tilde\rho
+
\tilde\nabla\cdot\mathbf v
\right)
+
\frac{\tilde{\mathbf J}\cdot\tilde{\mathbf E}}
{\tilde\rho},
\\
\frac{\partial\mathbf v}{\partial\tilde\tau}
&=
-\mathbf v\cdot(\tilde\nabla\mathbf v)
\nonumber\\
&\quad
+\mathbf v
\left[
\frac13
\left(
\mathbf v\cdot\tilde\nabla\ln\tilde\rho
+
\tilde\nabla\cdot\mathbf v
\right)
-
\frac{\tilde{\mathbf J}\cdot\tilde{\mathbf E}}{\tilde\rho}
\right]
\nonumber\\
&\quad
-\frac14
\tilde\nabla\ln\tilde\rho
+
\frac{\tilde{\mathbf J}\times\tilde{\mathbf B}}
{4\tilde\rho/3}
\nonumber\\
&\quad
+
\tilde\nu
\left[
\tilde\nabla^2\mathbf v
+
\frac13
\tilde\nabla
(\tilde\nabla\cdot\mathbf v)
\right],
\end{align}
where
\begin{equation}
\hat g
=
\frac{\tilde g\, w^2}{\phi_0^2\omega_*^2}\simeq0.137,
\qquad
\tilde g=\phi_0g_{\phi\gamma\gamma},
\end{equation}
and $\tilde{\mathbf J}\equiv\tilde\nabla\times\tilde{\mathbf B}$.

The electric field is obtained by inverting the generalized Ohm's law with the displacement current neglected,
\begin{equation}
\tilde{\mathbf E}
=
\frac{\tilde\sigma\mathbf K
+\tilde g\,\tilde\nabla\phi\times\mathbf K
+(\tilde g^2/\tilde\sigma)(\mathbf K\cdot\tilde\nabla\phi)\tilde\nabla\phi}
{\tilde\sigma^2+\tilde g^2|\tilde\nabla\phi|^2},
\end{equation}
with $\mathbf K=\tilde\nabla\times\tilde{\mathbf B}-\tilde\sigma\,\mathbf v\times\tilde{\mathbf B}+\tilde g\phi'\tilde{\mathbf B}$. Expanding to first order in $\tilde\eta=1/\tilde\sigma$ at fixed $\mathbf v$ and $\nabla\phi$ gives
\begin{equation}
\begin{aligned}
\tilde{\mathbf E}
&\simeq
-\mathbf v\times\tilde{\mathbf B}
\\
&\quad
+
\tilde\eta\Bigl[
\tilde\nabla\times\tilde{\mathbf B}
+
\tilde g\phi'\tilde{\mathbf B}
\\
&\qquad\qquad
-
\tilde g\,\tilde\nabla\phi\times(\mathbf v\times\tilde{\mathbf B})
\Bigr]
+O(\tilde\eta^2),
\end{aligned}
\end{equation}
and
\begin{equation}
\begin{aligned}
\tilde{\mathbf E}\cdot \tilde{\mathbf B}
&\simeq
\tilde\eta\Bigl\{
\tilde{\mathbf B} \cdot(\tilde\nabla\times\tilde{\mathbf B})
\\
&\qquad
+
\tilde g\bigl[(\phi'+\mathbf v\cdot\tilde\nabla\phi)\tilde B^2
\\
&\qquad\qquad
-(\mathbf v\cdot\tilde{\mathbf B})(\tilde{\mathbf B}\cdot\tilde\nabla\phi)\bigr]
\Bigr\}
+O(\tilde\eta^2).
\end{aligned}
\end{equation}
The simpler forms that omit the $O(\tilde g\,\mathbf v)$ gradient--velocity pieces hold when those contributions are neglected, when $|\nabla\phi|$ is small, or when $\mathbf v\approx0$; a high-conductivity limit alone does not drop them relative to the retained $\tilde\eta\,\tilde g\phi'\tilde{\mathbf B}$ term. With vanishing initial velocity the leading $\mathcal{R}_i$ estimate used in the main text remains applicable at $\tilde\tau_i$.
The cosmological background is not evolved as a dynamical variable. The functions $R(T)$ and $H(T)$ are built from the simulation's tabulated $g_*(T)$ fit, the conformal time follows from
\begin{equation}
\tau(T)
=
\tau_i
+
\int_{T_i}^{T}
\frac{1}{R^2(T)H(T)}\frac{dR}{dT}
\,dT,
\end{equation}
and the solver interpolates $R(\tilde\tau)$ and $\tilde{\mathcal H}(\tilde\tau)$ at each step. The same Wantz--Shellard fit is used in the background density and entropy-conservation scale-factor prescription. At $T_i=220$~MeV, $g_*(T_i)\simeq61.3$ fixes the units of Sec.~\ref{sec:dimless} and gives $\tilde m_\phi=0.647$.

The equations are solved with the Dedalus pseudospectral framework \cite{2020PhRvR...2b3068B}. Spatial derivatives are computed in Fourier space, while nonlinear terms are evaluated using $3/2$ grid padding before projection onto the retained modes. The MHD and pure-scalar evolutions use the second-order semi-implicit backward differentiation formula (SBDF2). The vacuum-Maxwell comparisons use the second-order Crank--Nicolson--Adams--Bashforth method (CNAB2) with a fixed step $\Delta\tilde\tau=0.005$.

The MHD runs use an adaptive CFL timestep based on the fluid and Alfv\'{e}n velocities, with safety factor $0.3$, initial step $\Delta\tilde\tau=0.001$ and maximum step $0.02$.
Periodic boundary conditions are imposed in all three directions.

Figure~\ref{fig:box_size} explains the box and spacing chosen in Sec.~\ref{sec:dimless}: several Hubble volumes must fit inside the box while the wall spans several grid cells. The dimensional comoving spacing is $\Delta x=L/N$, and its physical value is $R\Delta x$.

\begin{figure}[!htbp]
    \centering
    \includegraphics[width=0.8\linewidth]{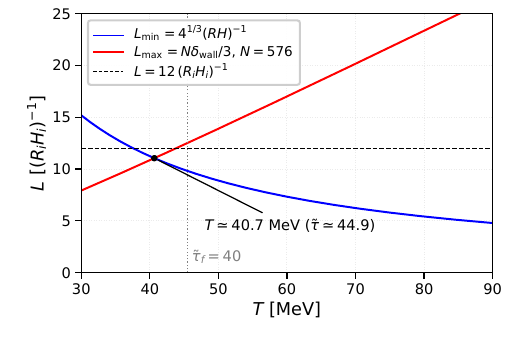}
    \caption{Box-size design in units of $(R_iH_i)^{-1}$. Blue: minimum length for four Hubble volumes; red: maximum length resolving the chosen wall width by three cells at $N=576$. Dashed: chosen length $12$; dotted: simulation endpoint. The curves use the prescribed simulation background and cross near $T\simeq40.7$~MeV.}
    \label{fig:box_size}
\end{figure}

For the physical kink $\phi_{\rm phys}=\phi_0\tanh(m_\phi z/2)$, the comoving width is $\delta_{\rm wall}=2/(Rm_\phi)$. Its lattice counterpart is $\tilde\delta_{\rm wall}=\omega_*\delta_{\rm wall}=2/(R\tilde m_\phi)$. At $\tilde\tau_f=40$, the simulation background gives $R\simeq7.90$ and $T\simeq45.5$~MeV. With $\tilde m_\phi=0.647$, the width is $\tilde\delta_{\rm wall}\approx0.391$, or $3.1\,\Delta\tilde x$.

The three-cell wall-width and four-Hubble-volume bounds are reached at $\tilde\tau\approx42$ and $49$, respectively, both beyond the simulation endpoint.

\endgroup

The endpoint width is close to the three-cell limit. We have not performed a resolution study or varied the scalar realization, so neither continuum convergence nor the scatter among realizations has been established. The formation bias and area deficit are measured at earlier times.

\section{Population bias, wall area and network lifetime}\label{app:area_response}

Figure~\ref{fig:area_parameter} compares the measured area histories with the reconstructions of Eq.~\eqref{eq:A_from_deltaA}.

A larger bias is accompanied by faster area loss. For the two largest-resistivity runs, over $18\le\tilde\tau\le40$ the areas fall from $0.670$ to $0.0835$ and from $0.819$ to $0.391$, respectively. At the smallest resistivity, the network and the pure-scalar control both retain $\mathcal A$ of order unity.

For the area parameter response we use $\mathcal A_s\equiv\mathcal A(\tilde\tau_A)$ at $\tilde\tau_A=18$, near the onset of scaling in the pure-scalar and weakly biased runs. Values at this time are linearly interpolated from the dense outputs.

\subsection{Initial area deficit}
\label{sec:res_area}
The reference-time deficit $\Delta\mathcal A_s$ is defined in Sec.~\ref{sec:vacuum_drive}. The four primary points with $\tilde\eta\ge5\times10^{-5}$ determine the capped power law in Eq.~\eqref{eq:deltaA_vs_Ri}, shown in Fig.~\ref{fig:deltaA18_vs_Ri}. The cap represents complete area loss; weak-drive points are excluded from this fit.

The weak-drive deficits are of order $10^{-3}$ and partly reflect differences in the timing of the area turnover at $\tilde\tau_A=18$. Unlike the unsigned bias, $\Delta\mathcal A_s$ has no positive lower bound by definition.

\begin{figure}[!htbp]
\centering
\includegraphics[width=0.8\linewidth]{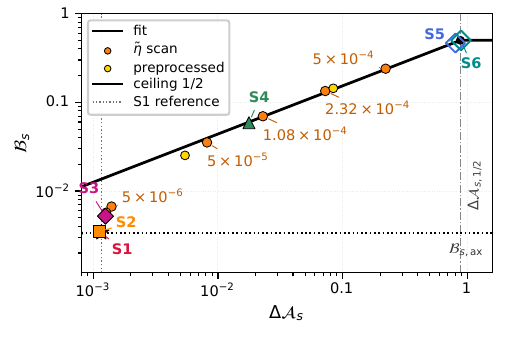}
\caption{Formation bias versus area deficit. Orange and gold circles denote runs without and with preprocessing; other markers denote S1--S6. The solid curve shows Eq.~\eqref{eq:Bs_from_deltaA_elim} with the extensions to the weak-drive and saturation limits. The dash-dotted line marks the saturation intersection; dotted guides mark the pure-scalar bias and S1 area deficit. The pure-scalar point lies outside the logarithmic axes.}
\label{fig:deltaBs_vs_deltaA}
\end{figure}

Eliminating $\mathcal R_i$ between the two calibrated power laws gives
\begin{equation}
\label{eq:Bs_from_deltaA_elim}
\mathcal{B}_s\simeq0.535\,(\Delta\mathcal{A}_s)^{0.543},
\end{equation}
for the unsaturated intermediate-drive points. Figure~\ref{fig:deltaBs_vs_deltaA} shows that stronger vacuum selection is accompanied by a larger area deficit already at formation. The extensions outside the fitted range are described in Sec.~\ref{app:aux_response_fits}.

Near the bias in the uncoupled run, the total bias need not follow the power law in the intermediate range because the seed can reinforce or oppose the induced signed bias (Sec.~\ref{app:calibration_checks}).

\subsection{Area evolution and network lifetime}
\label{sec:res_network}

\paragraph{Area evolution.}
After $\tilde{\tau}_A$ we track the area deficit relative to the pure-scalar history,
$\Delta\mathcal{A}(\tilde\tau)\equiv\mathcal{A}_{\mathrm{ax}}(\tilde\tau)-\mathcal{A}(\tilde\tau)$,
with a model formation-stage amplitude $\Delta A_0$.
Writing $r_A=1-\Delta A_0/\mathcal A_{s,\rm ax}$, Eq.~\eqref{eq:A_from_deltaA} can be written as
\begin{equation}
\label{eq:deltaA_grow}
\begin{aligned}
\frac{\mathcal A(\tilde\tau)}{\mathcal A_{\rm ax}(\tilde\tau)}
&=r_A\exp[-K_{\rm pop}(\tilde\tau-\tilde\tau_A)^{p_A}],\\
K_{\rm pop}&=-\kappa_A\ln r_A .
\end{aligned}
\end{equation}
The shape parameters are common to all five runs.

The coefficients $\kappa_A=0.04435$ and $p_A=1.686$ minimize the squared area residuals of the five primary histories over $18.05<\tilde\tau\le40$, with initial deficits fixed by Eq.~\eqref{eq:deltaA_vs_Ri}. The initial deficit of the weakest run is extrapolated from the fit. For $0<r_A<1$ the area approaches zero asymptotically, so radiation estimates use a fractional-area endpoint.

In this fit, the later suppression depends on the area deficit at formation.\footnote{Within this fitting ansatz, $\Delta A_0=0$ gives $\mathcal A=\mathcal A_{\rm ax}$; this identity is not a general dynamical constraint.} A persistent vacuum-pressure bias can instead suppress a network with initially matched areas if it does not alter formation-stage vacuum selection; Sec.~\ref{app:potential_comparison} compares the fitting forms.

\paragraph{Population-biased lifetime.}
To compare lifetimes, we fit the minority fraction with Eq.~(24) of Barbini et al.~\cite{Barbini:2026edx}, $F(\tilde\tau)=F(\tilde\tau_B)\exp[-x^{m+\alpha x}]$, where $x=(\tilde\tau-\tilde\tau_B)/\Delta\tilde\tau_{\rm ann}$ and $m$ and $\alpha$ are template parameters. The fitted $\Delta\tilde\tau_{\rm ann}$ is a characteristic decay timescale: at $\tilde\tau_{\rm ann}=\tilde\tau_B+\Delta\tilde\tau_{\rm ann}$, the template gives $F(\tilde\tau_{\rm ann})=F(\tilde\tau_B)/e$. The finite-volume fit criterion is given in Sec.~\ref{app:calibration_checks}.

As discussed in Sec.~\ref{sec:vacuum_drive}, only two fitted timescales fall within the simulated interval, so we fix the exponent of Ref.~\cite{Barbini:2026edx} rather than fitting it independently. Sec.~\ref{app:calibration_checks} discusses the reference-time dependence and finite-volume effects.

\subsection{Finite-volume effects and validity of the fits}
\label{app:calibration_checks}

With signed bias $\beta=(f_+-f_-)/2$, the induced change is $\Delta\beta=\beta-\beta_{\rm pure}$, whereas $\Delta\mathcal B_s=|\beta|-|\beta_{\rm pure}|$ can decrease for an opposed seed. The secondary fit $\Delta\mathcal B_s\simeq4.86\times10^3\mathcal R_i^{0.934}$ is used for the threshold conversion.

The runs in Table~\ref{tab:Ri_runs} cover $\mathcal R_i=9.80\times10^{-10}$--$9.80\times10^{-4}$. S1--S2 lie within $1.5$--$5\%$ of the bias in the uncoupled run, S3 and S4 give $0.87$ and $1.13$ times the bias fit, and S5 approaches saturation with $\mathcal B_s=0.467$ at $\mathcal R_i=6.19\times10^{-5}$, below the extrapolated intersection $\mathcal R_{\rm sat}^{B}=7.05\times10^{-5}$. S6 reaches $1/2$ by selecting one vacuum before a mature wall network forms. For S1, $V(0)/\rho_{{\rm rad},i}=m_\phi^2\phi_0^2/(8\rho_{{\rm rad},i})\simeq3.7$. This run tests the response on a prescribed background and does not represent a self-consistent radiation-dominated universe.

The primary runs share the same scalar and magnetic realizations and use $k_c=15$; S1--S6 use $k_c\simeq17$--$22$. The fits use the resistivity scan (four points for the initial area deficit), not the preprocessed runs or S1--S6. No fixed-$\mathcal R_i$ wavenumber scan, initial-field ensemble or resolution study was performed. Extrapolations to weak driving and saturation are illustrative.

For weak population bias in the quasi-scaling regime, Ref.~\cite{Barbini:2026edx} uses
\begin{equation}
\label{eq:B_power}
\mathcal{B}(\tilde{\tau})=\mathcal{B}_s\left(\frac{\tilde{\tau}}{\tilde{\tau}_B}\right)^{m},
\end{equation}
where $m$ is the growth exponent. The minority-fraction fit in Sec.~\ref{sec:res_network} uses times after $\tilde\tau_B$ with $F>1/N_H$, where $N_H=[12RH/(R_iH_i)]^3$ is the number of Hubble volumes in the box; the fitted scale does not directly measure disappearance. Our lifetime comparison uses $N_H\simeq160$, $90$ and $7$ at $\tilde\tau=15$, $18$ and $40$, respectively. Measuring the bias at $18$ instead of $15$ gives agreement with the Barbini prefactor at roughly the $20\%$ level, but changes the reference-time normalization. The $\tilde\eta=1.08\times10^{-4}$ lifetime scale lies just beyond $40$; smaller resistivities require longer extrapolations. S1--S6 are not included in this lifetime comparison.

\paragraph{Weak driving and saturation.}
\label{app:aux_response_fits}
The saturation intersections, $\mathcal R_{\rm sat}^{B}\simeq7.05\times10^{-5}$ and $\mathcal R_{\rm sat}^{A}\simeq7.10\times10^{-5}$, are fit extrapolations rather than measured critical thresholds. Figure~\ref{fig:deltaBs_vs_deltaA} displays the relation in the fitted range with extensions to the pure-scalar bias and the saturation limit. The saturation limit is reached at $\Delta\mathcal A_{1/2}\approx0.882$, obtained by evaluating the area fit at $\mathcal R_{\rm sat}^B$. The horizontal line marks the bias in the uncoupled run; the vertical guide marks the S1 area deficit.

\begin{figure}[!htp]
\centering
\subfloat[Coupling switched off\label{fig:control_off}]{\includegraphics[width=0.8\linewidth]{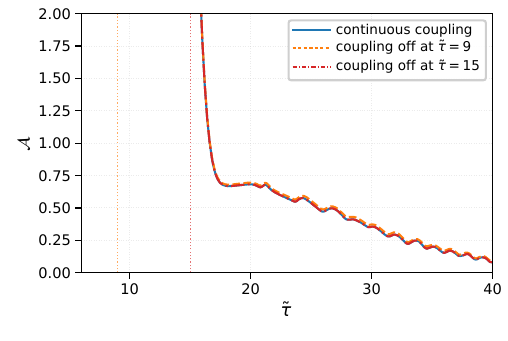}}
\hfill
\subfloat[Coupling switched on after formation\label{fig:control_on}]{\includegraphics[width=0.8\linewidth]{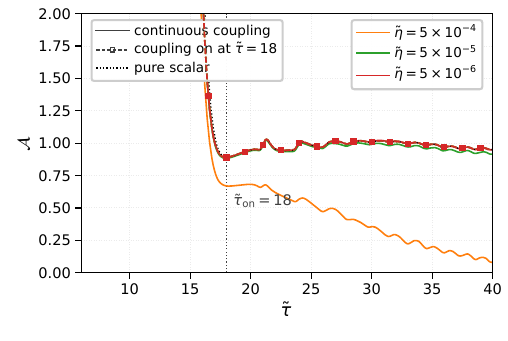}}
\caption{Area histories for different coupling intervals. (a) Full coupling compared with removal at $\tilde\tau=9$ or $15$, for $\tilde\eta=5\times10^{-4}$. (b) Coupling enabled only at $\tilde\tau=18$ gives nearly the same area evolution as the uncoupled scalar at all three resistivities, unlike continuous coupling.}
\label{fig:coupling_controls}
\end{figure}

\section{Timing of the electromagnetic coupling}
\label{app:supp_figs}

During the initial roll, the electromagnetic source shifts the scalar distribution in a direction determined by the magnetic helicity. This shift generates unequal vacuum occupation before substantial oscillations about the minima. In the primary runs, the source is strongest at $\tilde\tau\sim6$--$10$. Higher conductivity suppresses its electric component parallel to $\mathbf B$, while larger resistivity also dissipates magnetic helicity faster. The source histories are shown in Fig.~\ref{fig:magnetic_diagnostics}. The helicity-reversal and nonhelical controls are discussed in Appendix~\ref{app:helicity}.

\begin{figure}[!htbp]
\centering
\includegraphics[width=0.8\linewidth]{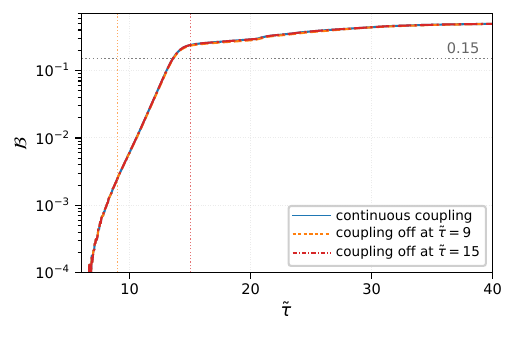}
\caption{Population bias for continuous coupling and coupling switched off at $\tilde\tau=9$ or $15$, for $\tilde\eta=5\times10^{-4}$. Vertical dotted lines mark the switch-off times. The complementary area histories appear in Fig.~\ref{fig:coupling_controls}(a).}
\label{fig:coupling_off_compare}
\end{figure}

Figures~\ref{fig:coupling_controls} and \ref{fig:coupling_off_compare} isolate when this source affects the network by comparing continuous coupling, removal during formation and activation after formation. The magnetic controls in Appendix~\ref{app:mhd} also show no appreciable scalar backreaction on magnetic energy, correlation length or source magnitude. Representative field distributions are given in Appendix~\ref{app:ndw2}.

For $\tilde\eta=5\times10^{-4}$, switching off the interaction at $\tilde\tau=15$ leaves the subsequent area evolution almost unchanged, whereas switching off at $9$ changes the area by up to about $12\%$ (Fig.~\ref{fig:coupling_controls}(a)). Figure~\ref{fig:coupling_off_compare} shows the corresponding occupation bias. Most of the bias is therefore generated before $\tilde\tau=15$.

\subsection{Driving after wall formation}
\label{app:delayed_coupling}

To test the electromagnetic tilt after wall formation, we compare full coupling, coupling enabled only at $\tilde\tau_{\rm on}=18$, and pure-scalar evolution at $\tilde\eta=5\times10^{-4}$, $5\times10^{-5}$ and $5\times10^{-6}$. The bare potential remains symmetric; the interaction can provide a local, time-dependent tilt as described in Sec.~\ref{sec:model}.

\begin{figure}[!htbp]
  \centering
  \includegraphics[width=0.8\linewidth]{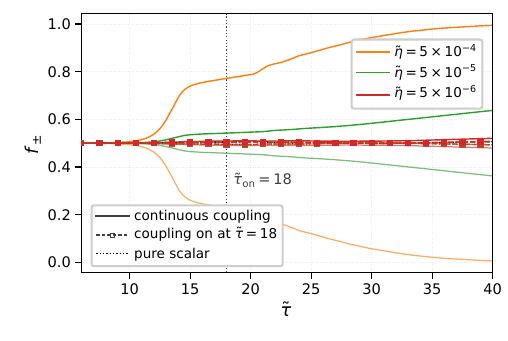}
  \caption{Vacuum fractions $f_\pm$ for the same continuously coupled, delayed-coupling ($\tilde\tau_{\mathrm{on}}=18$), and pure-scalar runs as in Fig.~\ref{fig:coupling_controls}(b). Vertical dotted line: $\tilde\tau=18$.}
  \label{fig:delayed_coupling_vacuum}
\end{figure}

In Fig.~\ref{fig:coupling_controls}(b) and Fig.~\ref{fig:delayed_coupling_vacuum}, the delayed-coupling area and occupation histories remain nearly identical to those of the pure-scalar controls through $\tilde\tau_f=40$. Together with the coupling-off tests (Fig.~\ref{fig:coupling_off_compare}), these results indicate that the bias generated during formation accounts for most of the later decay. The instantaneous tilt need not vanish even when no effect on the area or occupation is resolved.

\begin{figure*}[!t]
\centering
    \subfloat[\label{fig:time_evolution_EB}]{\includegraphics[width=0.27\linewidth]{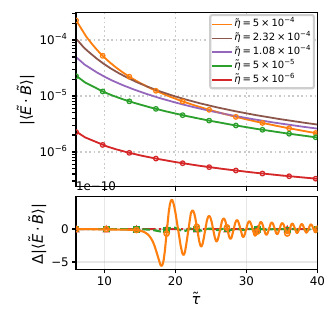}}
    %\hfill
    \subfloat[\label{fig:rhoB_evolution}]{\includegraphics[width=0.27\linewidth]{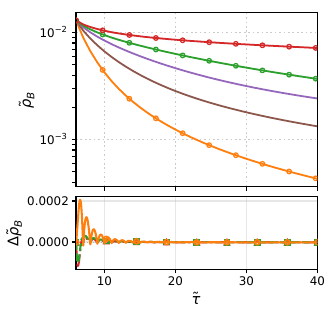}}
    %\hfill
    \subfloat[\label{fig:time_evolution_corr}]{\includegraphics[width=0.27\linewidth]{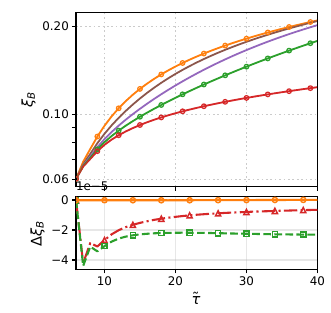}}
    \caption{Magnetic diagnostics: (a) source magnitude $|\langle\tilde{\mathbf E}\cdot\tilde{\mathbf B}\rangle|$, (b) comoving magnetic energy density $\tilde\rho_B$, and (c) correlation length $\xi_B$. The color key in (a) applies throughout. Solid curves denote scalar--MHD runs; open markers denote pure-MHD controls. Lower panels show $\Delta Q=Q_{\rm scalar\text{-}MHD}-Q_{\rm pure\,MHD}$ for the corresponding quantity.}
    \label{fig:magnetic_diagnostics}
\end{figure*}

\begin{figure*}[!t]
\centering
    \includegraphics[width=0.8\linewidth]{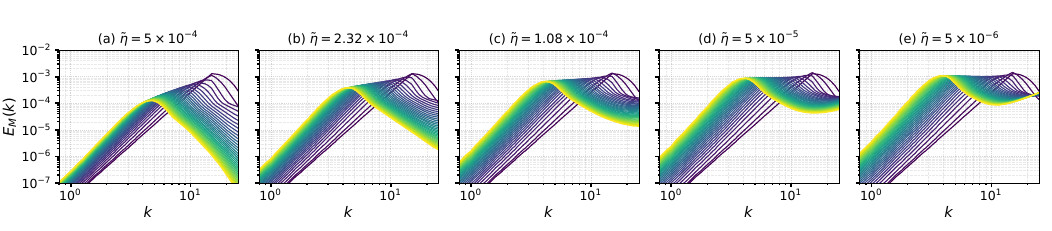}
    \caption{Magnetic energy spectra for the five resistivities, colored by conformal time from $\tilde\tau\simeq6$ (dark) to $40$ (yellow).}
    \label{fig:spectra_evolution_comparison}
\end{figure*}

\begin{figure*}[!t]
\centering
    \includegraphics[width=0.85\linewidth]{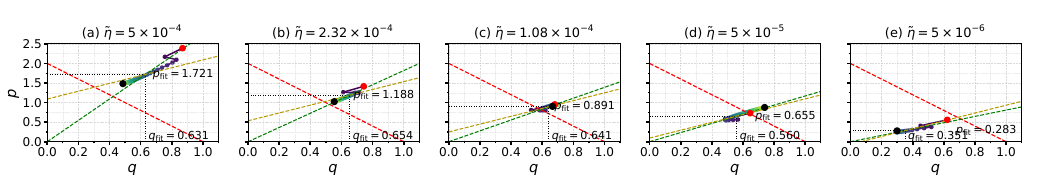}
    \caption{Magnetic decay trajectories in the $p$--$q$ plane, with exponents defined in Sec.~\ref{app:mhd}. Red and black dots mark the first and last finite-difference estimates; the red dashed line is $p=2(1-q)$. Colors progress from dark to yellow with increasing time. Green and yellow dashed lines show $p=(p_{\rm fit}/q_{\rm fit})q$ and $p-q=p_{\rm fit}-q_{\rm fit}$, respectively; dotted guides mark the fitted exponents.}
    \label{fig:p_q_comparison}
\end{figure*}

\subsection{Comparison with potential-biased decay}
\label{app:potential_comparison}

Population bias is an imbalance in vacuum occupation; it can drive network decay even when the vacuum energies are equal. Potential bias instead denotes a vacuum-energy difference, $\Delta V\ne0$, which exerts volume pressure on the walls~\cite{Larsson:1996sp,Correia:2018bias}. The two effects can coexist. Here the electromagnetic source can tilt the nonperiodic double well while active (Sec.~\ref{sec:model}), but the coupling controls show that the later area loss is governed predominantly by the occupation imbalance established during formation.

An exponential suppression of the comoving area density, $A_c/V_c\propto\tilde\tau^{-1}\exp[-K_{\rm pot}\tilde\tau^{p_{\rm pot}}]$, fits particular potential-biased simulations~\cite{Larsson:1996sp,Correia:2018bias}; it is not a universal decay law. Relative to the unbiased history, a comparison normalized to equal areas at $\tilde\tau_A$ is
\begin{equation}
\begin{aligned}
\mathcal A_{\rm pot}(\tilde\tau)
&\simeq\mathcal A_{\rm ax}(\tilde\tau)\\
&\quad\times\exp\!\left[-K_{\rm pot}
\left(\tilde\tau^{p_{\rm pot}}-\tilde\tau_A^{p_{\rm pot}}\right)\right],
\end{aligned}
\label{eq:area_potential_compare}
\end{equation}
with $K_{\rm pot}>0$. Equation~\eqref{eq:deltaA_grow} belongs to a similar family of exponential suppressions, but uses elapsed time $\tilde\tau-\tilde\tau_A$ and ties $K_{\rm pop}$ to the formation deficit through $r_A$. Thus $r_A=1$ removes subsequent suppression in our empirical law, whereas the potential-biased comparison allows equal reference areas with an independent $K_{\rm pot}>0$. The value $p_{\rm pot}=2$ applies to the two-dimensional matter-era simulations of Ref.~\cite{Correia:2018bias}, not universally. 

% Declare the Appendix E floats early for REVTeX placement.

Three-dimensional radiation-era simulations also exhibit similar early areas followed by potential-bias suppression~\cite{Ferreira:2024collapse}. Their fitted exponent $\nu_F\simeq3$ describes false-vacuum volume, not wall area. A potential bias can drive late decay even with little initial area deficit, although a tilt may also affect vacuum selection during formation. The fitted exponents alone do not distinguish the two mechanisms.

\section{Magnetic field evolution}
\label{app:mhd}
\label{app:spectra_pq}

Figure~\ref{fig:magnetic_diagnostics} compares the volume-averaged source, normalized comoving magnetic energy and correlation length. The pure-MHD and scalar--MHD histories nearly coincide, with no appreciable scalar backreaction. Angle brackets denote spatial averages over the periodic box.

Panel (a) plots $|\langle\tilde{\mathbf E}\cdot\tilde{\mathbf B}\rangle|$; the signed mean is negative for the negative-helicity runs. Periodic boundaries give $\mathrm d\tilde h/\mathrm d\tilde\tau=-2\langle\tilde{\mathbf E}\cdot\tilde{\mathbf B}\rangle$, with $\tilde h=\langle\tilde{\mathbf A}\cdot\tilde{\mathbf B}\rangle$ and $\tilde\nabla\times\tilde{\mathbf A}=\tilde{\mathbf B}$. The plotted magnitude is thus half the decay rate of $|\tilde h|$. The fractional initial helicity used to label the controls is defined in Sec.~\ref{app:helicity}.

The initial source ranges from $2.26\times10^{-4}$ at $\tilde\eta=5\times10^{-4}$ to $2.26\times10^{-6}$ at $5\times10^{-6}$. The largest-resistivity source decays fastest, crossing the intermediate curves around $\tilde\tau=10$--$17$. Starting from $\tilde\rho_{B,i}=0.013$, the magnetic energies end near $(4.3\times10^{-4},1.3\times10^{-3},2.4\times10^{-3},3.7\times10^{-3},7.2\times10^{-3})$, ordered by decreasing resistivity. For these initial fields, larger resistivity gives a stronger but shorter-lived source.

The full shell-integrated magnetic spectrum obeys $\tilde\rho_B=\int_0^\infty E_M(k)\,\mathrm dk$. The stored spectra and correlation-length diagnostic retain spherical shells only up to $k_{\rm Nyq}$. Accordingly, the quantity denoted by $\xi_B$ in the diagnostic plots is the cutoff estimator $\xi_B=\tilde\rho_B^{-1}\int_0^{k_{\rm Nyq}}k^{-1}E_M(k)\,\mathrm dk$, with the total lattice magnetic energy in the denominator. Here $k$ and $\xi_B$ are in the comoving lattice units of Sec.~\ref{sec:dimless}. This estimator omits the corner-mode contribution to the numerator; the plotted exponent $q$ characterizes this diagnostic rather than a full-spectrum correlation length. The spectra in Fig.~\ref{fig:spectra_evolution_comparison} shift toward smaller $k$, consistent with growing $\xi_B$. Preferential short-wavelength damping can produce this shift; it does not alone measure an inverse energy flux.\footnote{The displayed spectra stop at $k_{\rm Nyq}\simeq25.1$. Cartesian corner modes with $|\mathbf k|>k_{\rm Nyq}$ are omitted from the display; their energy contribution is modest initially but reaches roughly one quarter of the total at late times in the highest-conductivity run.}

The local exponents in Fig.~\ref{fig:p_q_comparison} are $p=-\mathrm d\ln\tilde\rho_B/\mathrm d\ln\tilde\tau$ and $q=\mathrm d\ln\xi_B/\mathrm d\ln\tilde\tau$, estimated between successive times. The annotations $p_{\rm fit}$ and $q_{\rm fit}$ come from log--log fits $\tilde\rho_B\propto\tilde\tau^{-p_{\rm fit}}$ and $\xi_B\propto\tilde\tau^{q_{\rm fit}}$ over $6\le\tilde\tau\le40$; they define the green and yellow guides. The red line $p=2(1-q)$ is a self-similar-decay reference \cite{Brandenburg:2016odr}, not an imposed fit.

\section{Effect of preprocessing the initial magnetic field}
\label{app:prep_vs_noprep}

The primary runs couple the initial helical random field to the scalar at $\tilde\tau_i=6$ without preprocessing. The alternative first evolves pure MHD and loads its magnetic field, velocity and density at the start of scalar evolution. All three preprocessed runs have $\tilde\rho_{B,i}\simeq0.013$, matching the primary magnetic-energy target. They share the primary scalar initial field at $\tilde\tau_i=6$, which remains fixed during MHD preprocessing.

\begin{figure}[!htp]
\centering
  \includegraphics[width=0.8\linewidth]{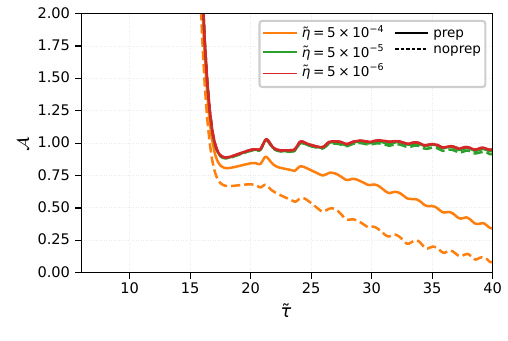}
  \caption{Area parameter with (solid) and without (dashed) MHD preprocessing. Colors distinguish the three resistivities. The vertical axis is truncated at $\mathcal A=2$.}
  \label{fig:prep_area}
\end{figure}

\begin{figure}[!htp]
\centering
  \includegraphics[width=0.8\linewidth]{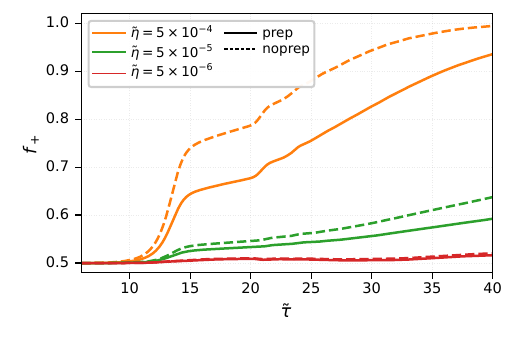}
  \caption{Vacuum volume fraction $f_+$ with and without preprocessing. Colors and line styles as in Fig.~\ref{fig:prep_area}. }
  \label{fig:prep_vacuum}
\end{figure}

Figures~\ref{fig:prep_area} and~\ref{fig:prep_vacuum} show the effect on walls. At $\tilde\eta=5\times10^{-6}$, the area histories nearly coincide; at $5\times10^{-5}$, they differ by about $2.5\%$ at $40$. At $5\times10^{-4}$, preprocessing raises the area from $0.670$ to $0.807$ at $18$ and from $0.0835$ to $0.343$ at $40$, while reducing the final positive-vacuum fraction to $0.936$. The bias still increases with resistivity, but prior MHD evolution changes its magnitude.

\par\medskip

\begin{figure}[!htp]
\centering
\includegraphics[width=0.8\linewidth]{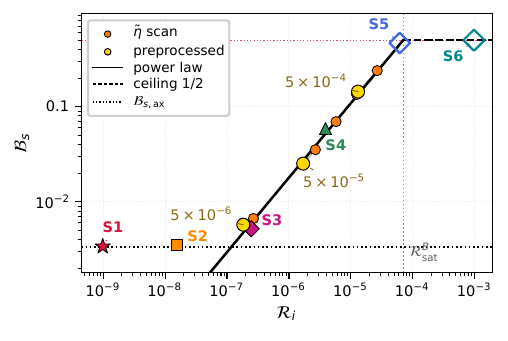}
\caption{Formation bias versus measured initial drive for preprocessed runs (gold circles). Other markers and the reference curve follow Fig.~\ref{fig:deltaBs15_vs_Ri}. Labels give the resistivity; gray segments connect runs with the same resistivity before and after preprocessing.}
\label{fig:deltaBs15_prep_measured}
\end{figure}

As shown in Fig.~\ref{fig:p_q_comparison}, the magnetic decay trajectories depend on the resistivity. For $\tilde\eta=5\times10^{-4}$, $5\times10^{-5}$ and $5\times10^{-6}$, MHD preprocessing shifts the spectral peaks from $15$ to approximately $7.3$, $9.5$ and $10.3$, respectively, while $\tilde\rho_{B,i}$ remains close to $0.013$. Using the measured spectral peaks in $\mathcal R_i=\hat g\tilde\eta(2\tilde\rho_{B,i}\tilde k_{*,i})$ reduces the estimated initial drive by about $30$--$50\%$. Figure~\ref{fig:deltaBs15_prep_measured} extends Fig.~\ref{fig:deltaBs15_vs_Ri} by including the three preprocessed runs. The resulting formation biases agree with the original power-law fit to within about $20\%$, suggesting that $\mathcal R_i$ also accounts for much of the dependence on the initial magnetic scale.

\section{Effect of the sign of the initial magnetic helicity}
\label{app:helicity}

We label the initial magnetic field by its fractional helicity~\cite{RoperPol:2022polarization},
\begin{equation}
\mathcal P_{M,i}=\frac{\int_0^\infty kH_M(k,\tilde\tau_i)\,\mathrm dk}
{2\int_0^\infty E_M(k,\tilde\tau_i)\,\mathrm dk},
\qquad -1\le\mathcal P_{M,i}\le1.
\end{equation}
Here $H_M$ is the magnetic-helicity spectrum, normalized by $\int H_M\,\mathrm dk=\langle\tilde{\mathbf A}\cdot\tilde{\mathbf B}\rangle$. The bound follows from $|kH_M|\le2E_M$. The controls use maximal negative helicity ($\mathcal P_{M,i}=-1$), zero helicity ($0$), and maximal positive helicity ($+1$). These labels specify the initial fields, rather than the helicity at every subsequent time.

\begin{figure}[!htp]
\centering
\includegraphics[width=0.8\linewidth]{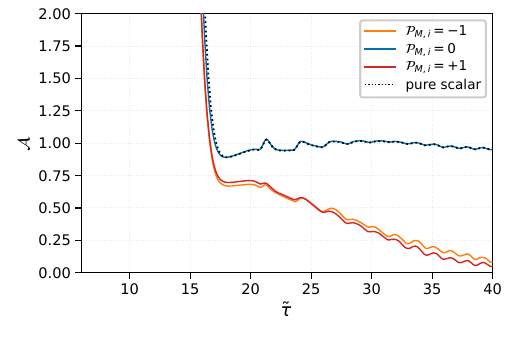}
  \caption{Area parameter for initial fractional helicities $\mathcal P_{M,i}=-1,0,+1$ at $\tilde\eta=5\times10^{-4}$, with the pure-scalar control (black dotted). The vertical axis is truncated at $\mathcal A=2$.}
  \label{fig:hel_area}
\end{figure}

\begin{figure}[!htp]
\centering
\includegraphics[width=0.8\linewidth]{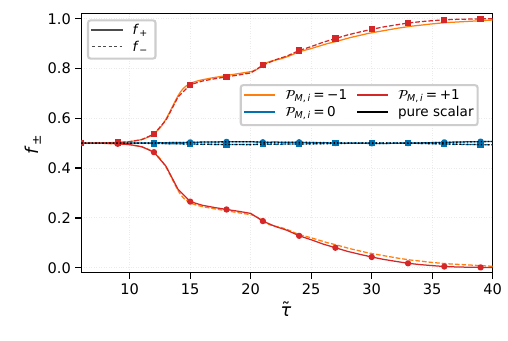}
  \caption{Vacuum fractions $f_+$ (solid) and $f_-$ (dashed) for the controls of Fig.~\ref{fig:hel_area}; the pure-scalar control is black dotted.}
  \label{fig:hel_vacuum}
\end{figure}

\begin{figure}[!htp]
\centering
\includegraphics[width=0.8\linewidth]{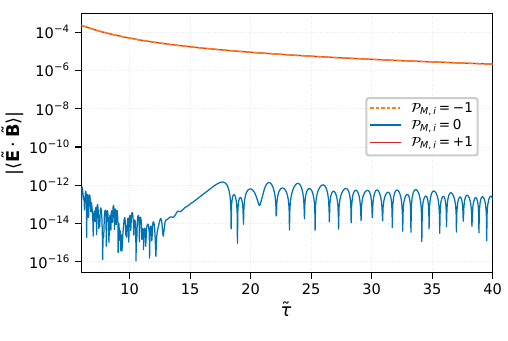}
\caption{Source magnitude $|\langle\tilde{\mathbf E}\cdot\tilde{\mathbf B}\rangle|$ for the magnetic controls of Fig.~\ref{fig:hel_area}. The helical magnitudes nearly coincide; the non-helical mean is much smaller.}
\label{fig:hel_EB}
\end{figure}

Figures~\ref{fig:hel_area}--\ref{fig:hel_EB} compare the non-preprocessed runs at $\tilde\eta=5\times10^{-4}$ for $\mathcal P_{M,i}=-1,0,+1$, together with the pure-scalar control. Reversing helicity reverses the favored vacuum, while the non-helical scalar evolution remains close to the pure-scalar result. The nonhelical field produces no appreciable additional bias in this realization (Fig.~\ref{fig:B_tau}). The two helical sources have nearly equal magnitudes and opposite signs, but the scalar histories are not exact mirrors for this fixed initial realization.

For a general spectrum, zero net magnetic helicity does not necessarily imply zero current helicity. The positive drive amplitude $\mathcal R_i$ in the main text does not encode this sign; a signed response law for general spectra has not been calibrated.

\FloatBarrier
\section{Cosmological constraints}
\label{sec:cosmology}

We use the fitted decay timescale to estimate the conditions for avoiding wall domination and ending wall-sourced gravitational-wave production before BBN. All physical conversions use the adopted simulation background, with $T_i=220$~MeV at $\tilde\tau_i=6$ and $g_*(T_i)\simeq61.3$. The template is $\Delta\tilde\tau_{\rm ann}/\tilde\tau_B\simeq0.2\,\mathcal{B}_s^{-0.8}$, with $\tilde\tau_B=15$ and $\tilde\tau_{\rm ann}=\tilde\tau_B+\Delta\tilde\tau_{\rm ann}$. Thresholds requiring evolution beyond the simulated interval rely on extrapolating this template.

The lifetime template gives a required total bias $\mathcal B_{\rm req}$. For the cosmological estimates we neglect the random imbalance of the finite simulation box and assume that the magnetic contribution $\mathcal B_{\rm mag}$ is approximated by the measured excess bias $\Delta\mathcal B_s=\mathcal B_s-\mathcal B_{s,\rm ax}$. The thresholds below therefore impose $\mathcal B_{\rm mag}\gtrsim\mathcal B_{\rm req}$ using this excess-bias fit. This prescription is an extrapolation from one initial realization; its dependence on the sign of the initial imbalance is discussed in Sec.~\ref{app:seed_thresholds}.

\subsection{Domain-wall domination and BBN}

\paragraph{Avoiding wall domination.}
We use the tension $\sigma_{\rm w}=\frac23m_\phi\phi_0^2$ and approximate the wall energy density by $\rho_{\rm DW}=2\mathcal{A}\sigma_{\rm w}H$.\footnote{This estimate neglects wall-velocity Lorentz factors, bending corrections and scalar radiation.} For the lifetime estimate, we approximate the product $\mathcal A\,\mathcal H\tau$ by its pure-scalar formation value $\mathcal A_{s,\rm ax}(\mathcal H\tau)_A$, so that $\mathcal A_{\rm ann}=\mathcal A_{s,\rm ax}(\mathcal H\tau)_A/(\mathcal H\tau)_{\rm ann}$. The fraction at annihilation is then
\[
\alpha_{\rm DW}=\frac{2\mathcal{A}_{s,\rm ax}(\mathcal H\tau)_A\sigma_{\rm w}}{3M_{\rm p}^2H_{\rm ann}(\mathcal H\tau)_{\rm ann}},
\]
where $M_{\rm p}$ is the reduced Planck mass and $(\mathcal H\tau)_A\simeq1.12$ at $\tilde\tau_A=18$. The factor $(\mathcal H\tau)_{\rm ann}$ is the simulation-background value, approximately $1.08$--$1.13$ over $\tilde\tau_B\le\tilde\tau\le40$. At the benchmark, $\phi_0=6\times10^{19}$~MeV, avoiding domination requires $\Delta\mathcal B_s\gtrsim0.00725$. The stronger condition $\alpha_{\rm DW}<0.1$ requires $\Delta\mathcal B_s\gtrsim0.037$. An unbiased network at this vacuum scale would dominate near $T\approx11$~MeV.

Table~\ref{tab:seed_thresholds} lists the two criteria and source amplitudes for this and two other vacuum scales.

Using $\Delta\mathcal{B}_s\simeq4856\,\mathcal{R}_i^{0.934}$, avoiding wall domination requires $\mathcal R_i\gtrsim5.8\times10^{-7}$, while $\alpha_{\rm DW}<0.1$ requires $\mathcal R_i\gtrsim3.3\times10^{-6}$. Both lie within the calibrated interval. The $\tilde\eta=5\times10^{-5}$ run, with $\Delta\mathcal{B}_s\approx0.032$, lies between the two thresholds. These illustrative bounds depend on the vacuum scale and the extrapolated lifetime template.

\paragraph{BBN and annihilation temperatures.}
For an illustrative BBN timing criterion, we require gravitational-wave production to end before $T\approx1$~MeV, adopting $\tilde\tau_{\rm gw}=\tilde\tau_B+2.85\Delta\tilde\tau_{\rm ann}$.\footnote{The factor $2.85$ is from the population-biased ensemble with $\mathcal B_s=0.081$ in Eq.~(35) of Ref.~\cite{Barbini:2026edx}; applying it to all runs is an additional assumption. This is a timing criterion and does not constrain the effects of wall-decay products on BBN.}

This criterion gives $\Delta\mathcal B_s\gtrsim0.0013$, independent of $\phi_0$ within the adopted template. It is weaker than the non-domination bounds and below the calibrated excess-bias range.

Applied to the total $\mathcal{B}_s$ of the five primary runs, the template gives $T_{\rm ann}\approx74$, $60$, $45$, $31$ and $10$~MeV in order of decreasing resistivity from $\tilde\eta=5\times10^{-4}$ to $5\times10^{-6}$. The last three times lie beyond $\tilde\tau_f=40$. The two template timescales fitted within the simulated interval instead correspond to approximately $68$ and $53$~MeV, consistent at the tens-of-percent level. This comparison is restricted to these two runs; smaller biases require extrapolation beyond the simulated evolution, and the physical temperatures depend on the assumed $T_i$.

\subsection{Effects of finite-volume population bias}
\label{app:seed_thresholds}
Table~\ref{tab:seed_thresholds} uses the excess bias $\Delta\mathcal B_s=|\beta|-|\beta_{\rm pure}|$ and the lifetime template and simulation background of Sec.~\ref{sec:cosmology}. The pure-scalar bias at $\tilde\tau_B=15$ is $b_0\simeq0.00335$. If an induced signed contribution of magnitude $q_{\rm ind}$ is superposed on this seed, the actual total bias is $\mathcal B_s=|q_{\rm ind}\pm b_0|$ for aligned or opposed signs. For a required total bias $\mathcal B_{\rm req}$, the aligned threshold is $q_{\rm ind}\ge\max(0,\mathcal B_{\rm req}-b_0)$, while the opposed high-bias branch requires $q_{\rm ind}\ge\mathcal B_{\rm req}+b_0$. When $b_0>\mathcal B_{\rm req}$, the opposed low-bias interval $0\le q_{\rm ind}\le b_0-\mathcal B_{\rm req}$ also satisfies the criterion. This illustrative superposition has not been tested for opposed seeds. The excess-bias diagnostic is defined relative to a single uncoupled run with the same initial field; an ensemble average would be needed to determine a cosmological mean response.

\begin{table}[tbp]
\centering
\small
\caption{Thresholds assuming negligible random population bias at formation, for avoiding wall domination and satisfying the BBN timing criterion of Sec.~\ref{sec:cosmology}. Source amplitudes $\mathcal R_i$ are in units of $10^{-7}$. $^\dagger$Below the calibrated range.}
\label{tab:seed_thresholds}
\setlength{\tabcolsep}{3pt}
\begin{tabular}{llcc}
\hline
$\phi_0$ [MeV] & Criterion & $\Delta\mathcal B_s$ & $\mathcal R_i$ \\
\hline
$6\times10^{19}$ & $\alpha_{\rm DW}<1$ & $0.00725$ & $5.81$ \\
 & $\alpha_{\rm DW}<0.1$ & $0.037$ & $33.2$ \\
$1.8\times10^{19}$ & $\alpha_{\rm DW}<1$ & $0.0016$ & $1.12^\dagger$ \\
 & $\alpha_{\rm DW}<0.1$ & $0.0068$ & $5.39$ \\
$9\times10^{19}$ & $\alpha_{\rm DW}<1$ & $0.0125$ & $10.4$ \\
 & $\alpha_{\rm DW}<0.1$ & $0.075$ & $70.5$ \\
any & BBN & $0.0013$ & $0.906^\dagger$ \\
\hline
\end{tabular}
\end{table}

\section{Gravitational waves and scalar relic abundance}
\label{sec:radiation}

We estimate gravitational waves with the population-bias template of Barbini et al.~\cite{Barbini:2026edx}, and wall-emitted scalar particles with an energy-release prescription adapted to our potential. We do not compute tensor perturbations or the emitted particle spectrum directly.

\subsection{Gravitational-wave spectrum}
We use the total measured $\mathcal B_s$ to set $\tau_{\rm ann}-\tau_B=0.2\tau_B\mathcal B_s^{-0.8}$ and the same $\tau_{\rm gw}=\tau_B+2.85(\tau_{\rm ann}-\tau_B)$ convention as in Sec.~\ref{sec:cosmology}. Define $\Omega_{\rm GW,0}(f)=\rho_{c,0}^{-1}d\rho_{\rm GW,0}/d\ln f$, where $f$ is today's frequency and $h=H_0/(100~\mathrm{km\,s^{-1}\,Mpc^{-1}})$. The present spectrum is $\Omega_{\rm GW,0}(f)h^2=\Omega_p h^2\mathcal S(f/f_p)$, with
\begin{equation}
\label{eq:gw_template}
\begin{aligned}
f_p&\simeq48\,{\rm nHz}\left(\frac{g_{*,\rm gw}}{100}\right)^{1/6}\frac{T_{\rm gw}}{150\,{\rm MeV}},\\
\Omega_p h^2&\simeq10^{-10}\epsilon_{\rm GW}
\left(\frac{g_{*,\rm gw}}{10.75}\right)^{-1/3}
\left(\frac{\alpha_{\rm ref,gw}}{0.01}\right)^2 .
\end{aligned}
\end{equation}
Here $g_{*,\rm gw}=g_*(T_{\rm gw})$, $H_{\rm gw}=H(T_{\rm gw})$, $\epsilon_{\rm GW}=0.06$, $\alpha_{\rm ref,gw}=2\sigma_{\rm w}/(3M_{\rm p}^2H_{\rm gw})$, and $\mathcal S(x)=4^{2.8}[x^{-3/2.8}+3x^{1/2.8}]^{-2.8}$, following Eqs.~(50)--(51) of Ref.~\cite{Barbini:2026edx}. Figure~\ref{fig:gw_population} displays the analytic template over the full frequency axes. The portion above $f/f_p=25$ is dashed to distinguish a formal high-frequency continuation from the range used for comparison with the numerical template. The asymptote scales as $f^{-1}$; a physical ultraviolet turnover, including finite wall-width effects, is not included in this continuation. The reference fraction in this normalization extrapolates a unit-area scaling network to $T_{\rm gw}$; it is not the residual wall fraction at that time.

The template describes radiation from the scalar walls only. In the late-decay, constant-$g_*$ limit, the fitted relation $\mathcal B_s\propto\mathcal R_i^{0.782}$ implies $f_p\propto\mathcal R_i^{0.626}$ and $\Omega_p h^2\propto\mathcal R_i^{-2.50}$; the finite reference-time shift and evolving $g_*$ modify these asymptotic powers.

\subsection{Scalar particle production}
\label{app:scalar_relic}
Adapting Ref.~\cite{Zhao:2025hzr} to our tension and area history gives the present density of stable particles emitted after $\tau_A$. With energy fraction $b_\phi$, number-weighted mean emission energy $\epsilon_{\rm em}m_\phi$ and conserved entropy,
\begin{equation}
\label{eq:scalar_relic_main}
\Omega_{\phi,\rm w}^{[A,d]}h^2=
\frac{s_0}{\rho_{c,0}/h^2}
\int_{t_A}^{t_d}\frac{b_\phi}{\epsilon_{\rm em}}\frac{Q_{\rm w}}{s}\,dt ,
\end{equation}
where $Q_{\rm w}$ is the wall energy released per physical volume and physical time, and $s$ is the entropy density. The energy-release approximation in terms of $\mathcal A$ is given below. The endpoint is defined by $\mathcal A(\tau_d)=0.01\mathcal A(\tau_A)$; it is separate from the GW-production endpoint $\tau_{\rm gw}$.

The reconstructed scalar-emission endpoints for the three largest drives are $\tilde\tau_d\simeq51$, $86$ and $151$, compared with GW endpoints of about $42$, $58$ and $87$. All require extrapolation beyond the simulation. The GW endpoints correspond to roughly $7\%$--$22\%$ of the reference area rather than the $1\%$ scalar-production criterion. The abundance bound in Fig.~\ref{fig:scalar_wall_abundance} excludes radiation before $\tau_A$ and depends on the emission prescription.

\paragraph{Mass, wall energy and particle number.}
Expanding Eq.~\eqref{eq:V_Z2_app} about either vacuum, $\phi=\pm\phi_0+\chi$, gives $V=\tfrac12m_\phi^2\chi^2+O(\chi^3)$, so the free small-amplitude excitations have mass $m_\phi$. A planar static wall has $\phi(z)=\phi_0\tanh(m_\phi z/2)$ and tension
\begin{equation}
\sigma_{\rm w}=\int_{-\phi_0}^{\phi_0}\sqrt{2V(\phi)}\,d\phi
=\frac23m_\phi\phi_0^2 .
\end{equation}
This tension, rather than that of a periodic axion potential, enters the abundance calculation. With our geometric area definition, the static-wall energy density is $\rho_{\rm w}=2\sigma_{\rm w}\mathcal A\mathcal H/R$.

If the emitted particles subsequently conserve their number, their physical number density satisfies
\begin{equation}
\dot n_\phi+3Hn_\phi=\frac{b_\phi Q_{\rm w}}{\epsilon_{\rm em}m_\phi}.
\end{equation}
Here $\epsilon_{\rm em}=\bar E_{\rm em}/m_\phi\ge1$ uses the ratio of emitted energy to emitted particle number. It cannot be determined from the area alone. Writing $Y_\phi=n_\phi/s$ and using $sR^3=\mathrm{const.}$ gives Eq.~\eqref{eq:scalar_relic_main}, with $s_0/(\rho_{c,0}/h^2)=2.744\times10^8~\mathrm{GeV}^{-1}$. The cancellation of $m_\phi$ between the number source and today's rest energy requires the constant mass used here; it does not assume that particles are nonrelativistic at emission.

\subsection{Relic abundance from the wall area}
To estimate the energy release from the area histories, we use $Q_{\rm w}=-\dot\rho_{\rm w}-H\rho_{\rm w}$, the static-wall energy-release approximation used in Ref.~\cite{Zhao:2025hzr}. It neglects wall-velocity corrections and continuing work on the wall component. The coupling controls motivate a weak late-time electromagnetic influence, but do not independently measure these corrections.

Define $u=\tilde\tau=\omega_*\tau$ and use $R_i=1$, $s_i=s(T_i)$ and $dt=R\,du/\omega_*$. Then
\begin{equation}
\label{eq:area_relic_integral}
\begin{aligned}
\Omega_{\phi,\rm w}^{[A,d]}h^2&=
\frac{s_0}{\rho_{c,0}/h^2}\frac{2\sigma_{\rm w}\omega_*}{s_i}
\frac{b_\phi}{\epsilon_{\rm em}}\,\mathcal J,\\
\mathcal J&=-\int_{u_A}^{u_d}R^2(u)
\frac{d}{du}\bigl[\mathcal A(u)\,\tilde{\mathcal H}(u)\bigr]du ,
\end{aligned}
\end{equation}
where $\tilde{\mathcal H}(u)=R(u)H(u)/\omega_*$, and $b_\phi$ and $\epsilon_{\rm em}$ are constant. We use the simulation background for $R(u)$ and the entropy, with $g_{*,s}=g_*$. At the benchmark, $H(t_A)/m_\phi\simeq0.029$.

To extrapolate the area, we use Eq.~\eqref{eq:A_from_deltaA} of Sec.~\ref{sec:vacuum_drive}, retaining $\kappa_A$, $p_A$ and the calibrated $\Delta A_0$ while replacing the measured pure-scalar history by $A_*=\mathcal A_{\rm ax}(u_A)=0.89133$. With $r_A=1-\Delta A_0/A_*$, this gives $\mathcal A_{\rm rec}(u)=A_*r_A^{1+\kappa_A(u-u_A)^{p_A}}$.
This extrapolation holds the pure-scalar area constant, unlike the reconstruction in Fig.~\ref{fig:area_parameter}. We integrate from $u_A=18$ to $\mathcal A_{\rm rec}(u_d)=0.01\mathcal A_{\rm rec}(u_A)$, giving $u_d=u_A+[\ln100/(-\kappa_A\ln r_A)]^{1/p_A}$. All three endpoints extrapolate the fitted tail beyond $u_f=40$. The two weaker primary drives are omitted because their longer area extrapolations are less constrained.

For the three drives in Fig.~\ref{fig:gw_population}, the benchmark gives
\begin{equation}
\label{eq:scalar_relic_numeric}
\Omega_{\phi,\rm w}^{[A,d]}h^2=C_i\frac{b_\phi}{\epsilon_{\rm em}}
\left(\frac{\phi_0}{6\times10^{16}\,{\rm GeV}}\right)^2 ,
\end{equation}
where $C_i\simeq(3.24,6.55,12.3)\times10^5$ at $m_\phi=2\times10^{-10}$~eV. Freezing the baseline at its final pure-scalar value instead raises these coefficients by about $12\%$, showing their dependence on the assumed pure-scalar area. The choice $\epsilon_{\rm em}=1.1$ in Fig.~\ref{fig:scalar_wall_abundance} is an illustrative input and is not inferred from a measured spectrum.

\subsection{Dependence on the vacuum scale and emission parameters}
At fixed mass, background and dimensionless response, $\Omega_{\phi,\rm w}^{[A,d]}\propto(b_\phi/\epsilon_{\rm em})\phi_0^2$, whereas $\Omega_p\propto\phi_0^4$. At the vacuum scale used in Fig.~\ref{fig:gw_population}, the estimated scalar abundance exceeds the observed dark matter abundance. A smaller scalar branching fraction, larger emission energy, particle decay or later entropy dilution would change this conclusion and must be specified in a particle-physics realization. Holding $\mathcal R_i$ and the dimensionless response fixed while changing $\phi_0$ requires compensating changes in the source parameters. We do not identify $\phi_0$ with a QCD-axion decay constant or impose a QCD mass--coupling relation.

\subsection{Mass dependence}
The simulations fix $m_\phi/\omega_*$ and the physical starting temperature. To extend Eq.~\eqref{eq:scalar_relic_numeric} to other masses, the main panel of Fig.~\ref{fig:scalar_wall_abundance} additionally assumes radiation-era similarity: $\omega_*\propto m_\phi$, $T_i\propto m_\phi^{1/2}$, and an unchanged dimensionless area history and background shape. Since $s_i\propto m_\phi^{3/2}$ and $\sigma_{\rm w}\propto m_\phi\phi_0^2$, this gives $\Omega_{\phi,\rm w}^{[A,d]}\propto\phi_0^2m_\phi^{1/2}$ and $\phi_{0,\max}\propto m_\phi^{-1/4}$ for fixed emission parameters. Changes of thermal degrees of freedom or source history can break this similarity. Only the marked benchmark mass is tied to the physical background simulated here.

\section{Observational backgrounds and model assumptions}
\label{sm:observational_context}

The GW comparison in Fig.~\ref{fig:gw_population} uses the corrected NANOGrav 15-year Hellings--Downs free-spectrum densities (version 2.0.0)~\cite{Agazie:2023GWB,Lamb:2023ceffyl,NANOGrav:2026KDE}. The displayed violins cover the central 95\% marginal interval in each of the lowest 14 frequency bins, with inner 68\% intervals. If $\rho_j$ is the timing-residual Fourier amplitude, $S_r(f_j)=T_{\rm obs}\rho_j^2$, and conversion to energy density uses $\Omega_{\rm GW,0}h^2=8\pi^4 f_j^5T_{\rm obs}\rho_j^2/H_{100}^2$, where $H_{100}=100~\mathrm{km\,s^{-1}\,Mpc^{-1}}$.

The EPTA and PPTA overlays are published fixed-index power-law amplitude intervals, not free-spectrum posterior densities. They assume $h_c=A(f/\mathrm{yr}^{-1})^{-2/3}$ and $\Omega_{\rm GW,0}h^2=2\pi^2f^2h_c^2/(3H_{100}^2)$. We use EPTA $\log_{10}A=-14.61^{+0.11}_{-0.12}$ (90\%)~\cite{Antoniadis:2023EPTA} and PPTA $A=(2.04^{+0.25}_{-0.22})\times10^{-15}$ (68\%)~\cite{Reardon:2023PPTA}. Their different statistical constructions are retained; these bands are not likelihood contours for our wall template or model-independent exclusion limits.

Projected power-law-integrated sensitivities for SKA, LISA, DECIGO and BBO use the tables accompanying Ref.~\cite{Schmitz:2020sensitivity}. The tabulated benchmarks have threshold signal-to-noise ratio one and observing times of 20 years (SKA) or one year (the interferometers). The Taiji and $\mu$Ares comparisons use the author-tabulated reference curves associated with Refs.~\cite{RoperPol:2022polarization,Sesana:2021muAres}. Their tabulated $\Omega_{\rm GW,0}$ values are multiplied by $h^2$ with $h=0.674$; the Schmitz tables already give $\Omega_{\rm GW,0}h^2$. The projected sensitivities use the assumptions of the respective studies and are not current exclusion bounds.

Figure~\ref{fig:scalar_wall_abundance} adds selected photon-coupling constraints, mapped to inverse vacuum scale for fixed $g_{\phi\gamma\gamma}\phi_0=10^{-3}$. The CAST boundary is the final-run $95\%$ limit $g_{\phi\gamma\gamma}<5.8\times10^{-11}~\mathrm{GeV}^{-1}$ for $m_\phi\lesssim0.02$~eV~\cite{CAST:2024final}. The other contours show globular-cluster evolution~\cite{Dolan:2022GC}, Chandra observations of H1821+643~\cite{SiskReynes:2021Chandra}, SN1987A gamma-ray conversion~\cite{Hoof:2023SN1987A}, and H.E.S.S. spectral irregularities~\cite{HESS:2013ALP}, using the contours collected in AxionLimits~\cite{OHare:2020AxionLimits}. They retain the stellar, supernova and magnetic-field assumptions of the respective analyses and are not combined statistically.

These limits constrain the photon coupling of small scalar excitations. They do not assume a QCD mass--decay-constant relation. Changing $g_{\phi\gamma\gamma}\phi_0$ shifts the limits on $1/\phi_0$. We omit bounds requiring additional couplings or a specified dark matter fraction. The dashed abundance boundaries assume the radiation-era similarity described in Appendix~\ref{sec:radiation}; only the dots correspond to the simulated mass.

\FloatBarrier
\section{Periodic complex-field model and field distributions}
\label{app:ndw2}

\subsection{Complex-field equations and simulation parameters}
The complex-field simulations use $\Phi=f_a(u_1+i u_2)$, with $\boldsymbol u=(u_1,u_2)$, $r_\Phi=|\boldsymbol u|$ and $\theta=\arg\Phi$. In the lattice units of Sec.~\ref{sec:dimless}, the dimensionless potential $\tilde V=V/(f_a^2\omega_*^2)$ is
\begin{equation}
\label{eq:complex_potential}
\begin{aligned}
\tilde V={}&\frac{\tilde\Lambda}{4}(r_\Phi^2-1)^2\\
&+\frac{\tilde m_a^2(T)}{N_{\rm DW}^2}
\bigl[1-r_\Phi\cos(N_{\rm DW}\theta)\bigr].
\end{aligned}
\end{equation}
We use the core-regularized potential of Ref.~\cite{Hiramatsu:2012sc}, Appendix~A.1, with $N_{\rm DW}=2$ and $\tilde\Lambda=0.2$. The vacuum fractions are $f_k=V_k/V_c$, assigning each grid point to the nearest angle $\theta_k=2\pi k/N_{\rm DW}$ modulo $2\pi$. Here $V_k$ is the assigned comoving volume and $V_c$ the box volume. For $N_{\rm DW}=2$, sector $k=0$ spans $-\pi/2\le\theta<\pi/2$ modulo $2\pi$ and $k=1$ its complement. Thus $f_0+f_1=1$, without amplitude weighting.

The two real components obey
\begin{equation}
\label{eq:complex_eom}
\begin{aligned}
\boldsymbol u''+2\tilde{\mathcal H}\boldsymbol u'
-\tilde\nabla^2\boldsymbol u
={}&-R^2\nabla_{\boldsymbol u}\tilde V\\
&+\frac{\hat g}{R^2}
\frac{\tilde{\mathbf E}\cdot\tilde{\mathbf B}}
{r_\Phi^2+\varepsilon}
\begin{pmatrix}u_2\\-u_1\end{pmatrix},
\end{aligned}
\end{equation}
where primes denote $\partial_{\tilde\tau}$ and $\nabla_{\boldsymbol u}$ differentiates with respect to the field components. Here $\tilde g=f_ag_{\phi\gamma\gamma}$ and $\hat g=\tilde g\,\rho_{{\rm rad},i}/(f_a^2\omega_*^2)$, as in the real-field normalization with $\phi_0$ replaced by $f_a$. The phase derivatives and angular source are regularized near string cores with $\varepsilon=10^{-20}$.

Unlike the constant-mass $Z_2$ scan, the complex-field mass scale increases as the background cools:
\begin{equation}
\label{eq:complex_mass_history}
\tilde m_a(T)=\tilde m_{a,\infty}
\min\!\left[1,\left(\frac{100\,{\rm MeV}}{T}\right)^{3.34}\right],
\end{equation}
with $\tilde m_{a,\infty}=\tilde m_\phi/2\simeq0.3234$. It starts at $\tilde m_a(T_i)\simeq0.023$ and saturates below $100$~MeV, whereas $\tilde m_\phi\simeq0.647$ is fixed throughout the $Z_2$ evolution. The choice $\tilde m_{a,\infty}=\tilde m_\phi/2$ matches the nominal late-time angular-wall width $1/(R\tilde m_{a,\infty})$ to the $Z_2$ width $2/(R\tilde m_\phi)$, permitting a comparable resolved evolution over $6\le\tilde\tau\lesssim40$. It does not equate the two mass histories. Both scalar-field models use the same prescribed Wantz--Shellard background~\cite{Wantz:2009cosmology}, $576^3$ grid and $\tilde L=72$.

The uncoupled, MHD-coupled and vacuum Maxwell complex-field runs share the same scalar initial realization.
% TODO: Verify and specify the initial phase and radial-field distributions, spatial correlations, and time derivatives from the complex-field simulation code.
The coupled runs use $\tilde g=10^{-3}$, $f_a=6\times10^{19}$~MeV and the same maximally negative-helicity initial magnetic field as the $Z_2$ comparisons, with $\tilde\rho_{B,i}=0.013$ and $\tilde k_c=15$. The MHD run uses $\tilde\eta=5\times10^{-4}$ and the Ohm inversion of Sec.~\ref{sec:comoving}, with the phase $\theta$ entering the electromagnetic coupling. The vacuum-Maxwell equations follow the axion--Maxwell formulation of Ref.~\cite{Zhao:2025hzr}, retaining the displacement current and omitting the conducting fluid, with the electric and magnetic sign conventions used here. The initial longitudinal electric field satisfies the scalar-modified Gauss constraint. The uncoupled run omits the electromagnetic source.

\onecolumngrid

\subsection{Field distributions in the periodic and nonperiodic models}

\begin{figure*}[!htbp]
    \centering
    \includegraphics[width=0.85\linewidth]{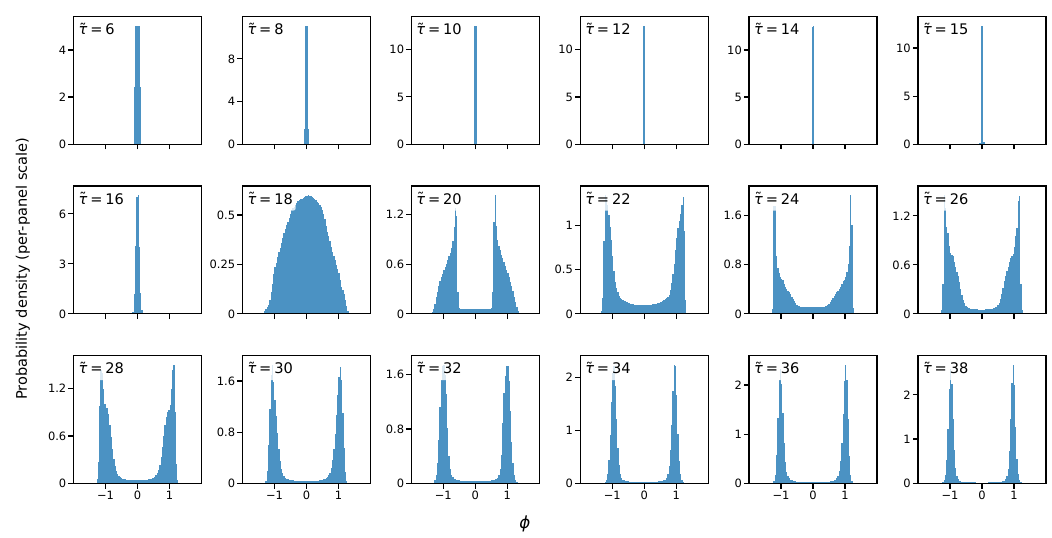}
    \caption{Unit-normalized volume probability densities of the lattice field $\phi$ in the pure-scalar run at the indicated $\tilde\tau$. Each panel has its own vertical scale. The field remains near the hilltop through $\tilde\tau\approx15$ and develops comparably populated vacuum peaks around $18$.}
    \label{fig:DW_formation_pureAxion}
\end{figure*}

\begin{figure*}[!htbp]
    \centering
    \includegraphics[width=0.85\linewidth]{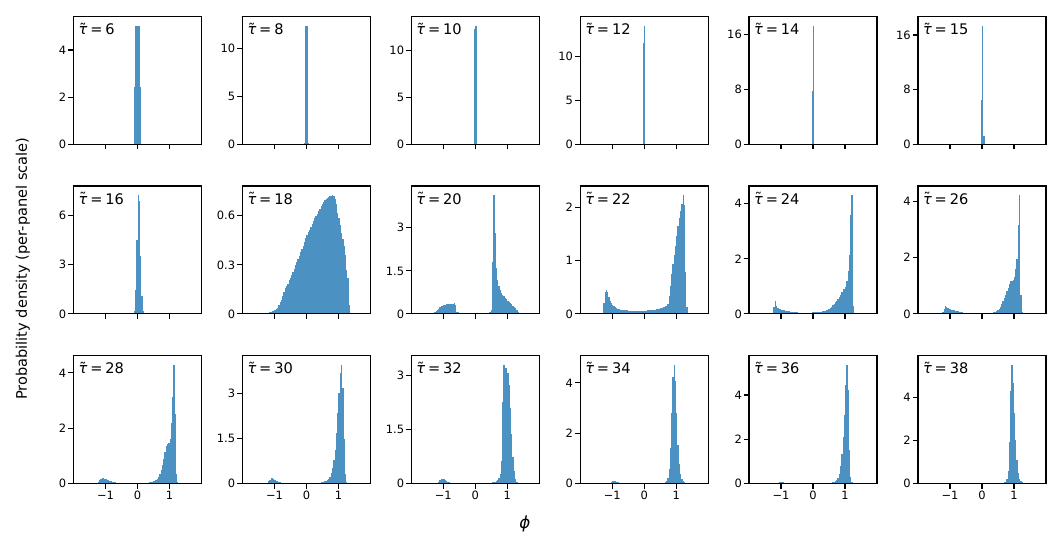}
    \caption{Volume histograms of $\phi$ for $\tilde{\eta}=5\times 10^{-4}$ ($\tilde{\sigma}=2\times 10^{3}$), as in Fig.~\ref{fig:DW_formation_pureAxion}. The minority peak is suppressed during formation and depleted by $\tilde\tau\approx30$.}
    \label{fig:DW_formation5e_4}
\end{figure*}

\begin{figure*}[!htbp]
    \centering
    \includegraphics[width=0.85\linewidth]{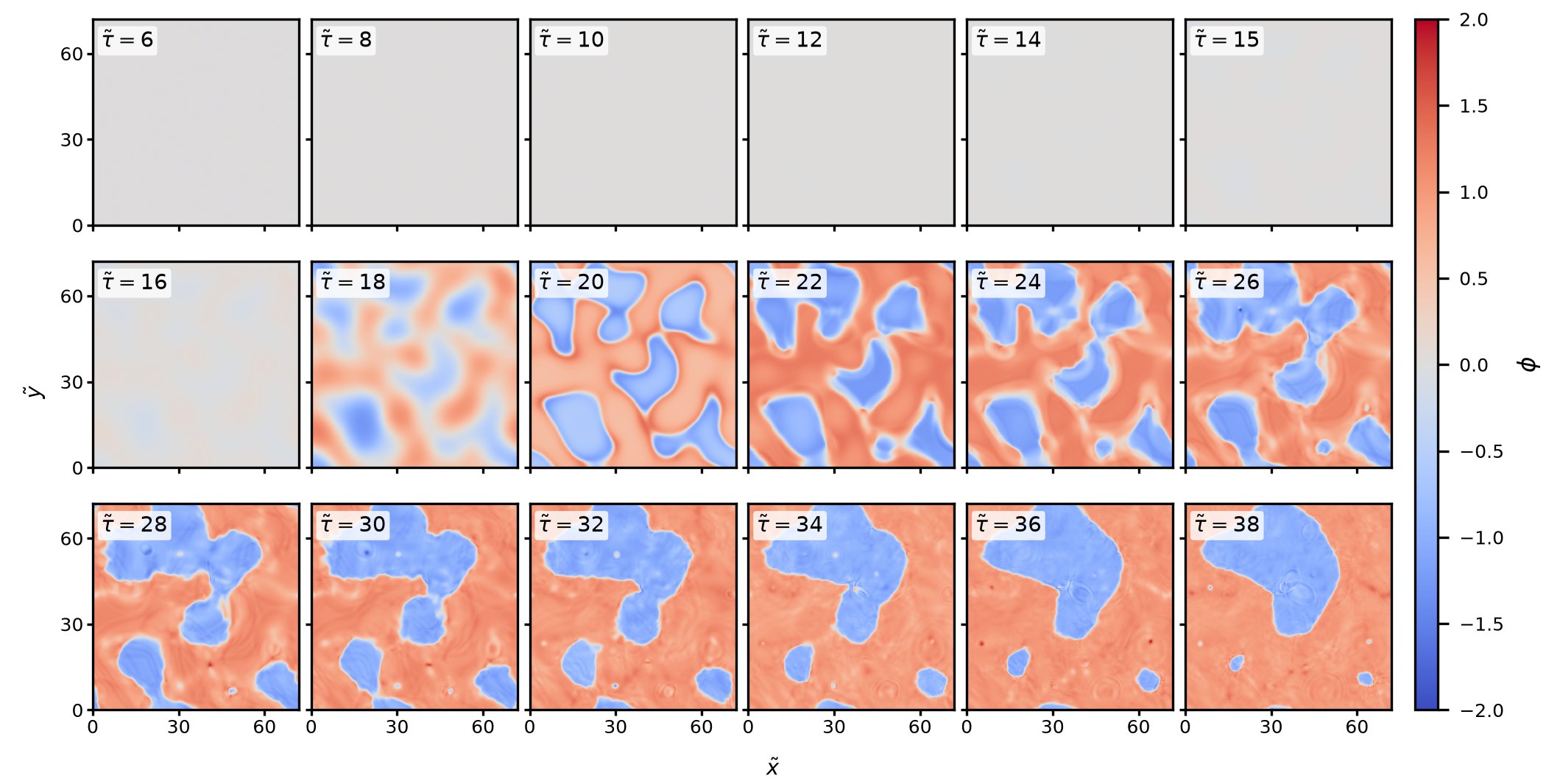}
    \caption{Two-dimensional slices of the pure-scalar field at the times indicated. Coordinates labeled ``code units'' are $\tilde x$ and $\tilde y$; color shows $\phi=\phi_{\rm phys}/\phi_0$.}
    \label{fig:DW_slice_pureAxion}
\end{figure*}

\begin{figure*}[!htp]
    \centering
    \includegraphics[width=0.85\linewidth]{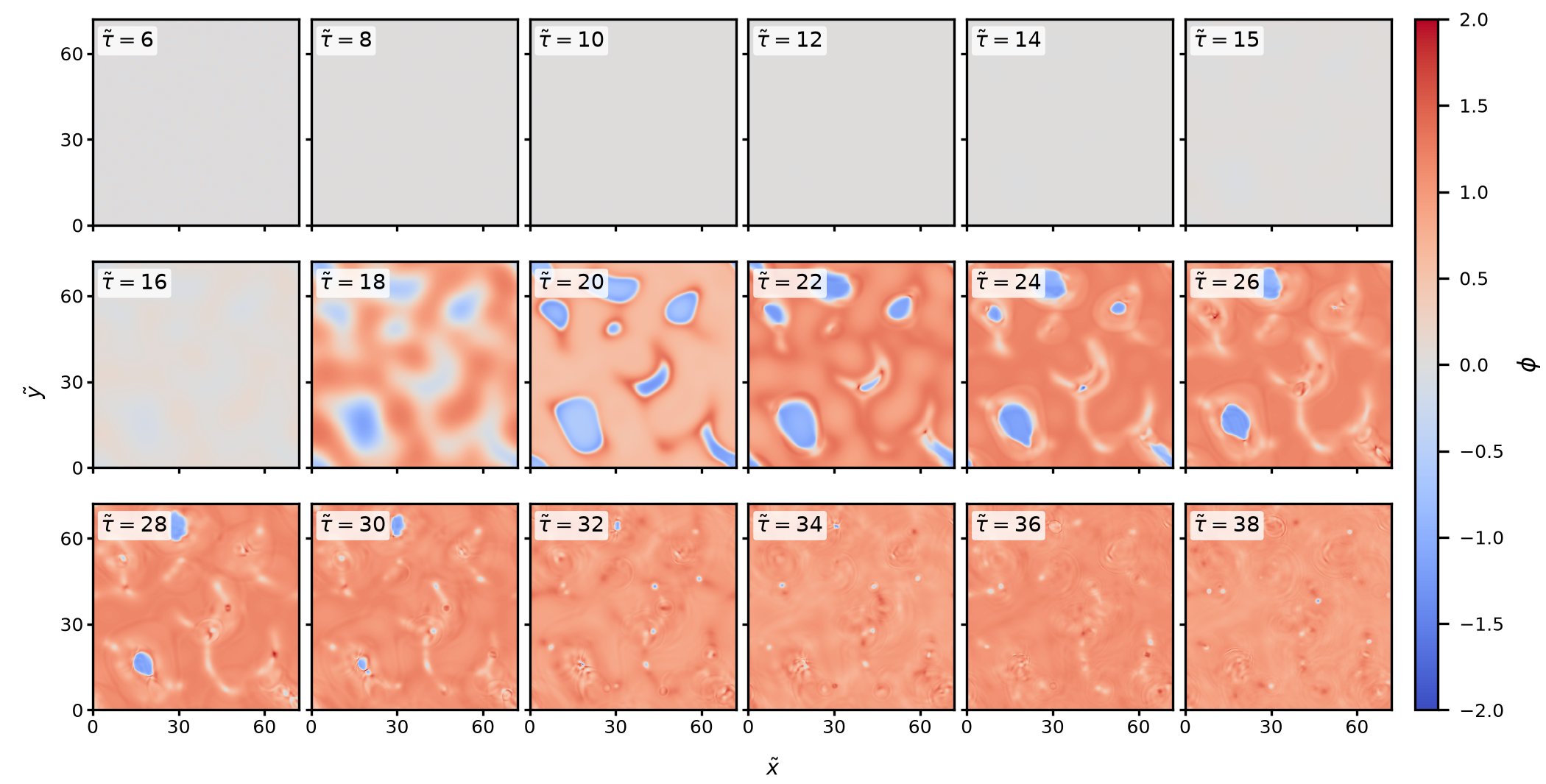}
    \caption{Two-dimensional slices of $\phi$ for $\tilde{\eta}=5\times 10^{-4}$ ($\tilde{\sigma}=2\times 10^{3}$), as in Fig.~\ref{fig:DW_slice_pureAxion}. The minority domains shrink as the favored vacuum expands.}
    \label{fig:DW_slice_5e_4}
\end{figure*}

Figures~\ref{fig:DW_formation_pureAxion}--\ref{fig:DW_slice_5e_4}
show the corresponding results for the nonperiodic scalar.
The minority vacuum is depleted in the MHD-coupled $Z_2$ model.

\begin{figure*}[!htbp]
    \centering
    \includegraphics[width=0.85\linewidth]{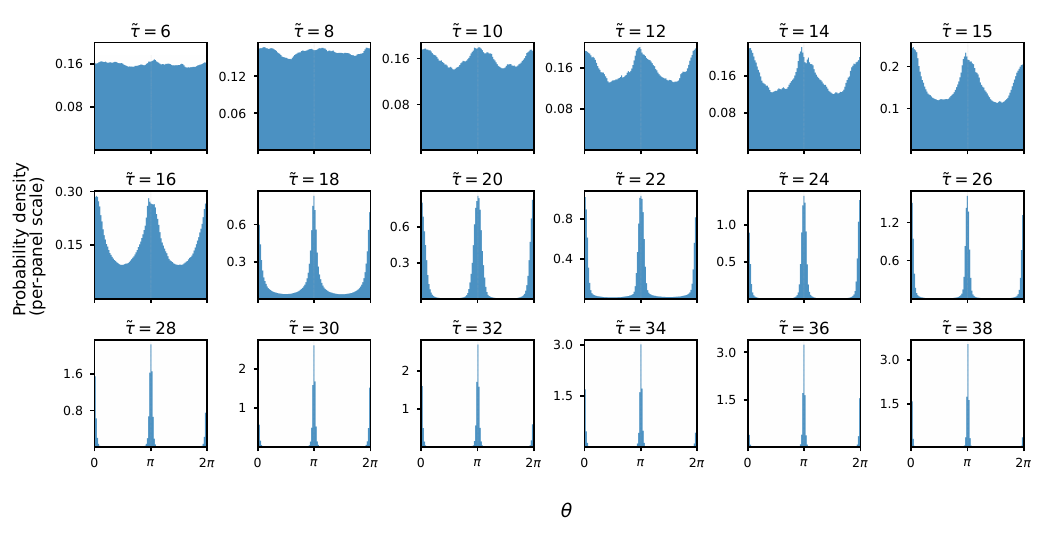}
    \caption{Volume histograms of $\theta=\arg\Phi$ for the $N_{\rm DW}=2$ pure complex-field run at the same 18 times as Fig.~\ref{fig:DW_formation_pureAxion}. The phase bins are centered at $\theta=0$; $0$ and $2\pi$ represent the same vacuum. Vertical scales are panel-dependent; occupation is given by $f_k$. }
    \label{fig:ndw2_pure_hist}
\end{figure*}

\begin{figure*}[!htbp]
    \centering
    \includegraphics[width=0.85\linewidth]{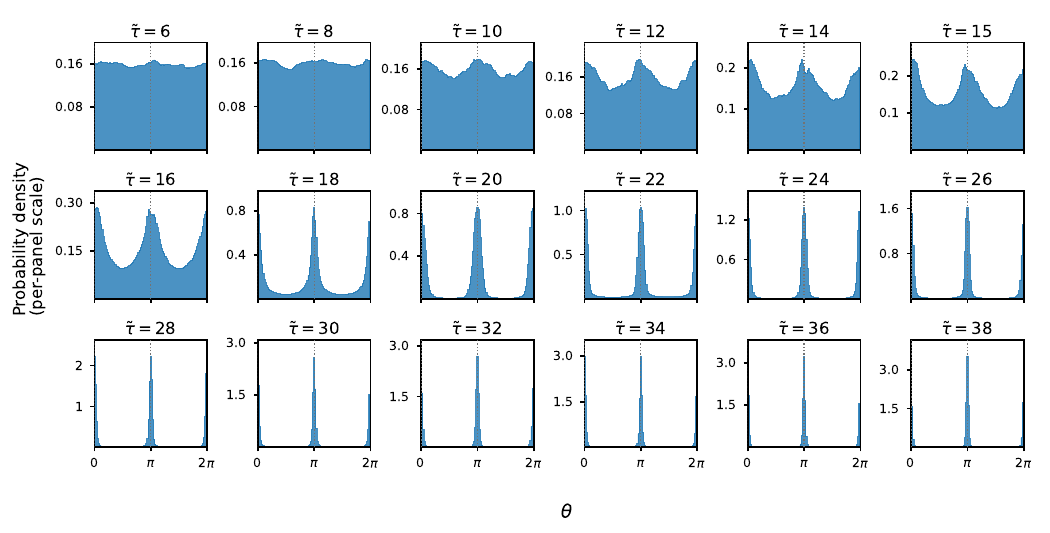}
    \caption{Phase histograms for the MHD-coupled complex field, with the binning and times of Fig.~\ref{fig:ndw2_pure_hist}.}
    \label{fig:ndw2_hist}
\end{figure*}

\begin{figure*}[!htbp]
    \centering
    \includegraphics[width=0.85\linewidth]{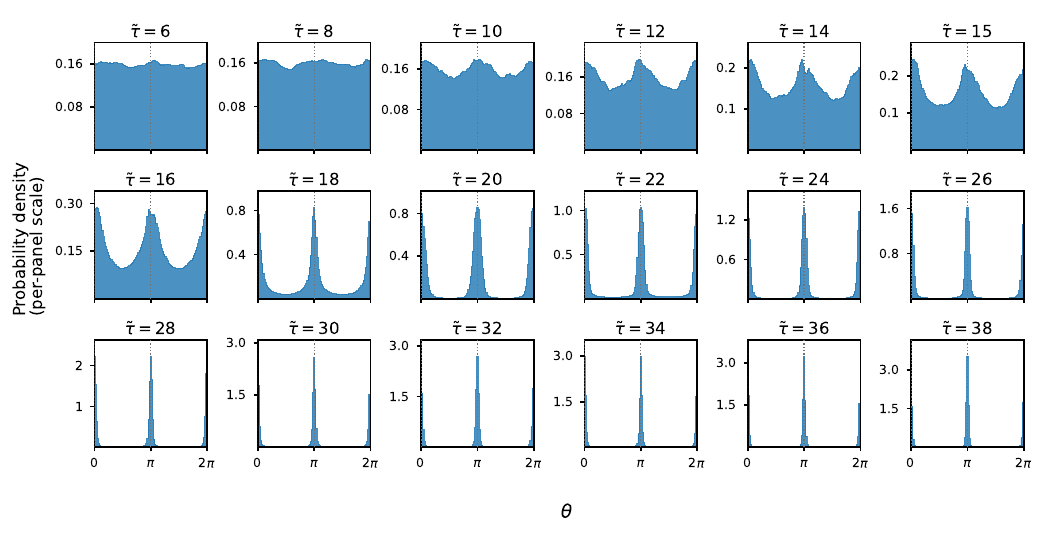}
    \caption{Phase histograms for the vacuum Maxwell-coupled complex field, with the binning and times of Fig.~\ref{fig:ndw2_pure_hist}.}
    \label{fig:ndw2_maxwell_hist}
\end{figure*}

\begin{figure*}[!htp]
    \centering
    \includegraphics[width=0.85\linewidth]{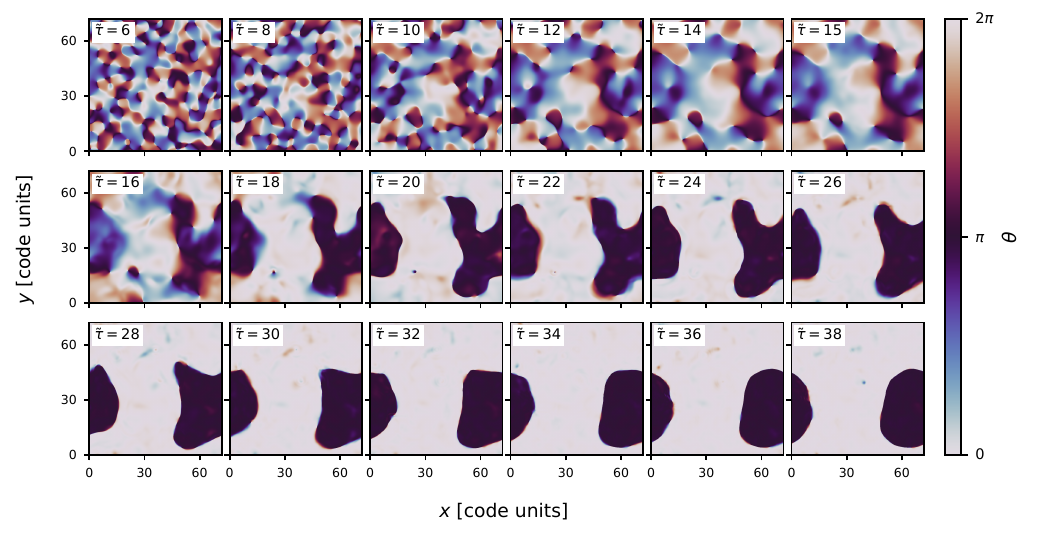}
    \caption{Midplane slices of $\theta$ for the $N_{\rm DW}=2$ pure complex-field run at the same 18 times as Fig.~\ref{fig:ndw2_pure_hist}. The cyclic phase colors identify $0$ and $2\pi$.}
    \label{fig:ndw2_pure_slice}
\end{figure*}

\begin{figure*}[!htp]
    \centering
    \includegraphics[width=0.85\linewidth]{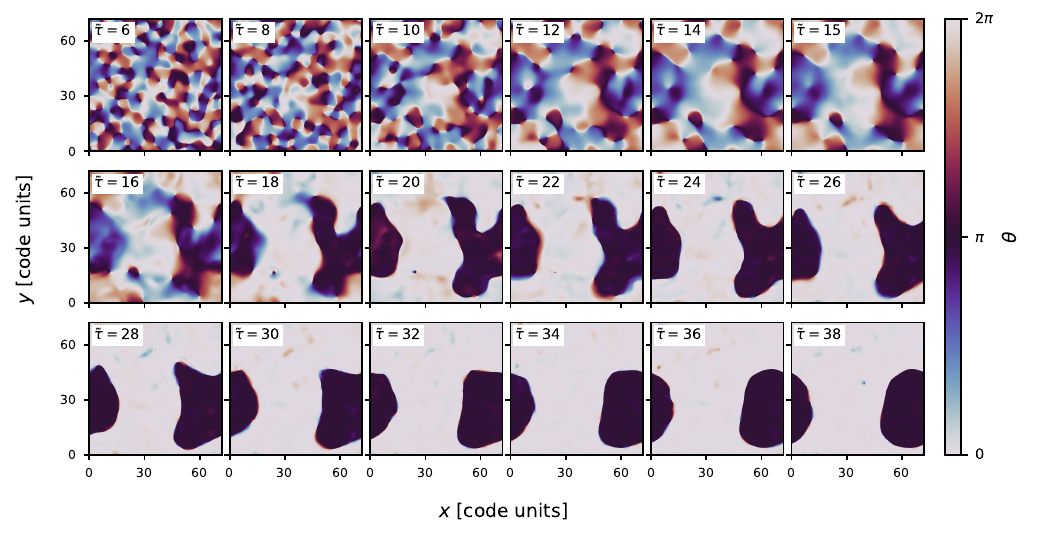}
    \caption{Same as Fig.~\ref{fig:ndw2_pure_slice} for the coupled MHD run. Both vacua of the periodic $N_{\rm DW}=2$ potential remain in the box.}
    \label{fig:ndw2_slice}
\end{figure*}

\begin{figure*}[!htbp]
    \centering
    \includegraphics[width=0.9\linewidth]{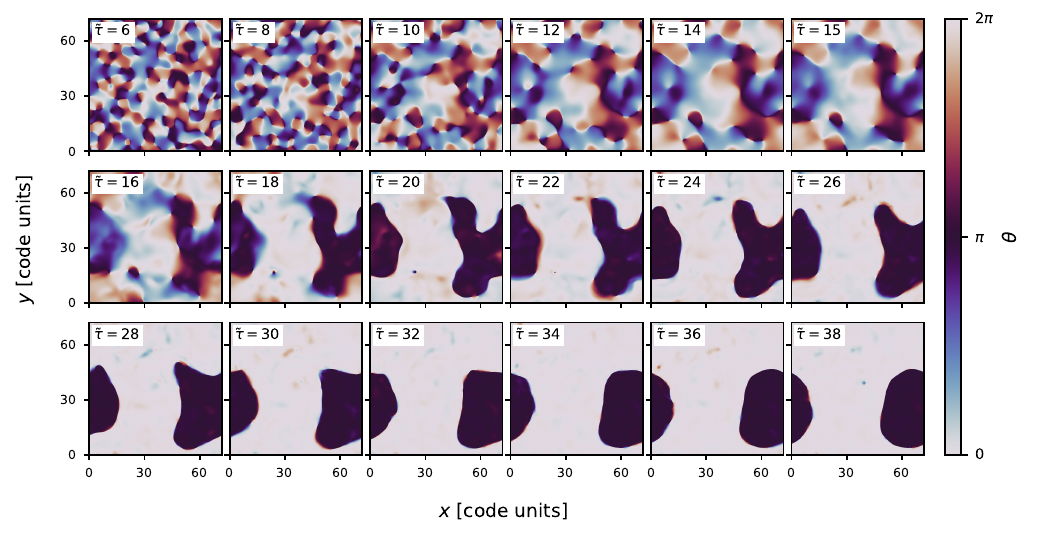}
    \caption{Same as Fig.~\ref{fig:ndw2_pure_slice} for the vacuum-Maxwell run, at the same 18 times as Fig.~\ref{fig:ndw2_maxwell_hist}. }
    \label{fig:ndw2_maxwell_slice}
\end{figure*}

The phase histograms and slices of the complex field are shown in Figs.~\ref{fig:ndw2_pure_hist}--\ref{fig:ndw2_maxwell_slice}. They sample 18 conformal times between $\tilde\tau=6$ and $38$.
The phase bins are centered at zero, with $\theta=0$ and $2\pi$ representing the same vacuum. Both vacuum peaks persist in the three complex-field runs, consistent with the occupation fractions in Sec.~\ref{sec:periodic_limit}. 

We assign each grid point to the nearer vacuum phase, $0$ or $\pi$ modulo $2\pi$, and denote the corresponding volume fractions by $f_0$ and $f_1$. At the end of the simulations, $\tilde\tau\simeq40$, the nearest-vacuum phase fractions are $(f_0,f_1)\simeq(0.495,0.505)$ in the uncoupled, MHD and vacuum Maxwell complex-field runs. Both vacua remain populated, with occupation histories close to those of the uncoupled complex field. For the periodic model of Appendix~\ref{app:ndw2} and its angular electromagnetic coupling, the transformation $\theta\to\theta+2\pi/N_{\rm DW}$ leaves the coupled equations invariant and cyclically permutes the vacua. If the initial statistical ensemble is invariant under this transformation, all vacuum fractions have the same expectation value. Since $\sum_k f_k=1$, it follows that $\langle f_k\rangle_{\rm ens}=1/N_{\rm DW}$, where $\langle\cdot\rangle_{\rm ens}$ denotes an ensemble average. This does not exclude an imbalance in an individual realization or changes in network dynamics that preserve the symmetry.

\FloatBarrier

\onecolumngrid

\clearpage
\twocolumngrid
\bibliographystyle{apsrev4-2}
\bibliography{references}
\end{document}